\documentclass[11pt,reqno,preprint]{article}

\usepackage{jheppub,epsfig,amssymb,amsmath,mathrsfs,hyperref,graphicx}
\usepackage{slashed}
\usepackage[shortlabels]{enumitem}
\usepackage{shuffle}
\usepackage{tikz-feynman}
\usepackage{multirow,bbm}
\usepackage[makeroom]{cancel}
\usepackage{caption}
\usepackage{subcaption}

\newcommand{\spa}[1]{\langle#1\rangle}
\newcommand{\spb}[1]{[#1]}
\newcommand{\tr}{\text{tr}}
\newcommand{\asym}{{\wedge^2F}}
\newcommand{\asyma}{{\wedge^2\bar{F}}}

\newcommand{\Li}{\text{Li}}
\newcommand{\e}{\epsilon}
\newcommand{\eps}{\epsilon}
\newcommand{\lam}{\lambda}
\newcommand{\mycomment}[1]{}
\def\be{\begin{equation}}
\def\ee{\end{equation}}
\def\ba{\begin{eqnarray}}
\def\ea{\end{eqnarray}}
\newcommand{\bea}{\begin{eqnarray}}
\newcommand{\eea}{\end{eqnarray}}
\newcommand{\nn}{\nonumber}

\title{\boldmath On gauge amplitudes first appearing at two loops}

\author[1]{Lance J. Dixon}
\author[1,2]{and Anthony Morales}

\affiliation[1]{SLAC National Accelerator Laboratory, Stanford University
	\\Stanford, CA 94309, USA}
\affiliation[2]{Physics Department, Stanford University
	\\Stanford, CA 94309, USA}

\emailAdd{lance@slac.stanford.edu}
\emailAdd{ammoral@stanford.edu}

\abstract{
  We study scattering amplitudes in massless non-abelian gauge theory
  where all outgoing gluons have positive helicity.  It has been argued
  recently by Costello that for a particular fermion representation
  (8 fundamentals plus one antisymmetric-tensor representation in $SU(N)$)
  the one-loop amplitudes vanish identically.  We show that this vanishing
  leads to previously-observed identities among one-loop color-ordered partial
  amplitudes.
  We then turn to two loops, where Costello has computed the all-plus
  amplitudes for this theory, as rational functions of the kinematics
  for any number of gluons using the celestial chiral algebra (CCA) bootstrap.
  We show that in dimensional regularization, these two-loop
  amplitudes are not rational, and they are not even finite as $\e\to0$.
  However, the finite remainder for four gluons agrees with the formula
  by Costello.
  In addition, we provide a mass regulator for the infrared-divergent
  loop integrals; with this regulator, the CCA bootstrap formula
  is recovered exactly.
  Finally, we use the CCA bootstrap to compute the double-trace terms
  in the theory at two loops for an arbitrary number of gluons.
}

\preprint{ \begin{flushright} SLAC--PUB--17784 \end{flushright}}

\dedicated{Dedicated to the memory of Stefano Catani}

\begin{document}
\maketitle
\flushbottom
	

\section{Introduction}
\label{sec:intro}

The study of scattering amplitudes has seen great advances in recent years.
On the more applied side, computing higher-point and higher-loop amplitudes
in the Standard Model has allowed for more precise comparisons to data
collected at particle colliders
(see e.g.~refs.~\cite{Heinrich:2020ybq,Andersen:2024czj} and
references therein).  On the more formal side, amplitudes are fascinating
theoretical objects in their own right.  They provide insight into the
behavior and symmetries of a theory, as well as exhibiting previously
unforeseen mathematical structures.  Having explicit analytic expressions
for amplitudes is paramount for finding such structures, and for better
understanding aspects of quantum field theory.
	
Often, direct calculation of amplitudes by evaluating Feynman diagrams
can be bypassed for more computationally efficient methods.
In particular, a general understanding of the singular behavior of
amplitudes can allow them to be ``bootstrapped" to higher orders in
perturbation theory, or for a greater number of scattering particles.
This program has had remarkable success in $\mathcal{N}=4$ supersymmetric
Yang-Mills in the planar limit (see
e.g.~refs.~\cite{Caron-Huot:2020bkp,Dixon:2022rse}
and references therein).

Amplitudes in ordinary, non-supersymmetric Yang-Mills (YM) theory
remain more challenging.  There have been remarkable recent advances in
computing the full-color all-helicity massless QCD
amplitudes for $2\to3$ scattering at two
loops~\cite{Agarwal:2023suw,DeLaurentis:2023nss,DeLaurentis:2023izi}
and for $2\to2$ scattering at three
loops~\cite{Caola:2021rqz,Caola:2021izf,Caola:2022dfa}.
These amplitudes have a rather intricate analytic structure, and pushing
directly to one more loop or one more leg may be difficult.

Another avenue for progress, which we will pursue here, is to investigate
the simplest possible helicity configuration, called ``all-plus'',
when all $n$ outgoing gluons have the same positive helicity.
Such amplitudes vanish for any $n$ in any supersymmetric massless gauge
theory~\cite{Grisaru:1976vm,Grisaru:1977px,Parke:1985pn},
and therefore they vanish at tree level in YM theory.
At one loop, in any massless gauge theory,
their unitarity cuts vanish in four dimensions,
and they are infrared (IR) and ultraviolet (UV) finite, rational
functions of the spinor products of the external momenta, which
are known for an arbitrary number of gluons~\cite{Mahlon:1993si,Bern:1993qk}.

Self-dual Yang-Mills theory~(sdYM)~\cite{Yang:1977zf} involves path integrals over only self-dual gauge field configurations. Classically, sdYM is integrable~\cite{Belavin:1978pa,Tze:1982gf,Chau:1982mn}. For free plane waves, such configurations include only the positive-helicity gluons. Interactions between plane waves include a $({-}{+}{+})$ vertex~\cite{Parkes:1992rz,Chalmers:1996rq}, but not the parity conjugate $({+}{-}{-})$ vertex.  At tree level, one can build the one-minus amplitude $({-}{+}{+}\cdots{+})$ by sewing together $({-}{+}{+})$ vertices, but this vanishes on shell. At loop level, the same sewing leads to the one-loop all-plus amplitudes~\cite{Cangemi:1996rx,Cangemi:1996pf}, which validates the suggestion that the non-vanishing of these amplitudes can be considered an anomaly in the conservation of the currents associated with integrability of sdYM~\cite{Bardeen:1995gk,Bittleston:2022nfr}.

At two loops, the connection to sdYM becomes less clear.
Two-loop all-plus gauge theory amplitudes were first computed for four
gluons using generalized unitarity~\cite{Bern:2000dn,Bern:2002tk}.
For five external gluons, the leading-color terms were computed first
numerically~\cite{Badger:2013gxa}, and later
analytically~\cite{Gehrmann:2015bfy,Dunbar:2016aux}.
The nonplanar integrands in the pure-glue theory were
found in ref.~\cite{Badger:2015lda}, and the complete nonplanar results
are available in refs.~\cite{Agarwal:2023suw,DeLaurentis:2023izi}.
For $n>5$, the polylogarithmic part of the leading-color result
was proposed for arbitrarily many gluons in ref.~\cite{Dunbar:2016cxp},
and the rational part was computed using an augmented recursion
relation for $n=6$~\cite{Dunbar:2016gjb} and $n=7$~\cite{Dunbar:2017nfy}.
(The planar $n=6$ integrand was presented in ref.~\cite{Badger:2016ozq}.)
Full-color results for $n=6$ in pure gauge theory were given in
ref.~\cite{Dalgleish:2020mof}.
The all-$n$ result for a particular color structure has been conjectured
in ref.~\cite{Dunbar:2020wdh}, and checked numerically for $n=8$ and 9
in refs.~\cite{Kosower:2022bfv,Kosower:2022iju}
(where the $n<8$ rational results were also checked).
Many of these results rely on $D$-dimensional generalized unitarity
for the construction of integrands, although the polylogarithmic results
in refs.~\cite{Dunbar:2016aux,Dunbar:2016cxp,Dunbar:2019fcq}
carry out the cuts four-dimensionally,
and the rational parts in
refs.~\cite{Dunbar:2016aux,Dunbar:2016gjb,Dunbar:2017nfy,Dunbar:2019fcq,
Dalgleish:2020mof,Dunbar:2020wdh} are constructed recursively.

The connection between twistors, string theory,
and tree-level gluon scattering amplitudes of (mostly)
positive helicity goes back to Nair~\cite{Nair:1988bq},
Witten~\cite{Witten:2003nn}, the MHV rules of Cachazo, Svr\v{c}ek and
Witten~\cite{Cachazo:2004kj}, and the derivation of these rules from
the YM action by Mason~\cite{Mason:2005zm}.  They have also been
derived from a twistor action~\cite{Boels:2007qn}.  These works
clarify the relations between sdYM and tree level amplitudes.
The MHV rules were applied to compute tree-level form factors of operators
composed of anti-self-dual field strengths,
e.g.~$\tr(F^2_{\rm ASD})$~\cite{Dixon:2004za}.
The all-plus and one-minus form factors for this operator were computed
at one loop in a non-supersymmetric $SU(N)$ theory in
ref.~\cite{Berger:2006sh}.

Recently, a novel bootstrap method for amplitudes in special theories
has been suggested in ref.~\cite{Costello:2022wso}.  It stems from a
combination of ideas from celestial holography, twisted holography, and
twistor theory.
In some sense, it is a loop level generalization of the earlier
tree-level work~\cite{Mason:2005zm,Boels:2007qn}.
In this method, the cancellation of an anomaly in a theory that lives
in twistor space allows for the existence of a chiral algebra,
the elements of which are in bijection with the states of the theory.
The correlators of the chiral algebra correspond to form factors of the
theory. The operator product expansions (OPEs) between the elements of
the chiral algebra are used to constrain the pole-structure of correlators,
the residues of these poles being lower-loop or lower-point correlators.
In this way, one can bootstrap the form factors of these theories.
	
In ref.~\cite{Costello:2023vyy}, this
\textit{celestial chiral algebra (CCA) bootstrap} was used to compute
a two-loop $n$-gluon all-plus-helicity form factor in sdYM
with Weyl fermions transforming in the representation
\begin{equation}
  R_0 \equiv 8F\oplus8\bar{F}\oplus\asym\oplus\asyma
\label{therep}
\ee
of the Lie algebra of $SU(N)$. Here $F$ is the fundamental representation,
and $\asym$ is the antisymmetric tensor representation.  In terms
of Dirac fermions, the representation has 8 fundamentals (quarks)
plus one antisymmetric tensor.  It solves the anomaly cancellation condition
from the six-dimensional twistor-space theory~\cite{Costello:2022wso},
\be
\tr_{R_0}(X^4) = \tr_G(X^4),
\label{eq:cancelanom}
\ee
for any generator $X$ of the $SU(N)$ Lie algebra, where $G$ denotes
the adjoint representation.  The form factor is for an operator
$\tfrac{1}{2} \tr(B\wedge B)$, involving an adjoint-valued,
antisymmetric, anti-self-dual tensor field $B_{\mu\nu}$,
which is used to enforce self-duality of the gauge field.

The sdYM form factor computed in ref.~\cite{Costello:2023vyy}
should reproduce scattering amplitudes in YM for arbitrary $n$.
Due to the anomaly cancellation condition, the one-loop amplitude
should vanish in this theory.  As we will see, this condition
implies identities among the QCD all-plus partial amplitudes.
The identities include the ``three-photon vanishing'' relations
first noticed in ref.~\cite{Bern:1993qk}.  A more general set of linear
relations was found in ref.~\cite{Bjerrum-Bohr:2011jrh};
we will show that these relations are all explained by the vanishing
of the one-loop all-plus amplitude for representation $R_0$.

The relevant two-loop sdYM form factor was computed for all $n$ in
ref.~\cite{Costello:2023vyy}.  The four-point result is
\begin{align}
\label{KCformula}
\begin{split}
  \mathcal{A}_{4,\text{sdYM}}^\text{2-loop}\ =\ 
		  \frac{g^6}{(4\pi)^4}\rho
  \bigg[& \bigg( 12 N - 4 \frac{s^2+4st+t^2}{st}
                - \frac{24}{N} \bigg)
		\big( \tr(1234) + \tr(1432) \big)
		\\
		&+ \bigg( 24 + \frac{24}{N} \bigg) \tr(12)\tr(34) \bigg]
		+ \mathcal{C}(234),
\end{split}
\end{align}
where
\be
\rho = i\frac{\spb{12}\spb{34}}{\spa{12}\spa{34}},
\ee
and $s=(k_1+k_2)^2$ and $t=(k_2+k_3)^2$ are the four-point
Mandelstam variables.  We use the shorthand notation
\be
  \tr_R(ij\cdots k) = \tr_R(t^{a_i}t^{a_j}\cdots t^{a_k}),
\ee
which is the trace over the generators $t^a$ of the Lie algebra of $SU(N)$
in an arbitrary representation $R$.
Throughout this paper, traces without a subscript,
as in eq.~\eqref{KCformula}, will mean the trace over
fundamental-representation generators.
The ``$+\mathcal{C}(234)$'' instructs one to add the two
non-trivial cyclic permutations of $(2,3,4)$ acting on the previous
expression.

In this paper, we wish to investigate the relation between the sdYM
form factor given in eq.~(\ref{KCformula}) and all-plus amplitudes in
ordinary YM.  The two-loop all-plus four-point amplitude in QCD
was computed in dimensional regularization in
refs.~\cite{Bern:2000dn,Bern:2002tk}.  Here we will replace the
fermion loops for QCD (i.e.~for fermions in the fundamental
(+ antifundamental) representation only) with fermion loops
in the representation $R_0$ in eq.~\eqref{therep}.
Then we can directly compare
the form factor in sdYM to the two-loop amplitude in YM.
The double-trace term is not provided in ref.~\cite{Costello:2023vyy},
so we compute it in Appendix \ref{sec:dtcomputation}.
Our results agree \textit{only after} UV renormalization and after
subtracting off the universal two-loop IR divergences given by
Catani~\cite{Catani:1998bh}.  This statement does not
disprove eq.~\eqref{KCformula}; rather,
the discrepancy most likely arises from
the fact that the CCA bootstrap technique keeps all momenta four-dimensional,
in contrast to dimensional regularization.  We resolve the discrepancy
by using a different IR regularization scheme, namely a mass regularization
of the loop integrands.  With this scheme, the two-loop
four-point sdYM form factor
equals the YM amplitude, and we suppose that the same will be true for
$n>4$.  We also argue that the $n$-gluon sdYM result
gives the finite remainder of the YM amplitude
in dimensional regularization.  This result could provide a check
of higher-point two-loop all-plus helicity amplitudes, once all the
fermionic and subleading-color terms become available.


\section{An overview of the CCA bootstrap}
\label{sec:CCAoverview}

In this section, we provide a non-rigorous overview of a method
used to bootstrap certain two-loop amplitudes~\cite{Costello:2023vyy}.
We will refer to this method as the
\textit{celestial chiral algebra (CCA) bootstrap}.
Positive- and negative-helicity states of sdYM on
twistor space are in one-to-one correspondence with local operators
in an (extended) chiral algebra. The conformal blocks of this algebra
are the local operators in the self-dual theory. Therefore, correlation
functions of the chiral algebra in a given conformal block correspond to
form factors of the gauge theory. Moreover, the OPEs in the algebra are
collinear limits of states in the field theory. This suggests that one
can use the chiral algebra to ``bootstrap'' form factors of sdYM
by using the analytic properties of the OPEs.
	
A requirement for the existence of a chiral algebra is the associativity
of its OPEs. Associativity fails at the first loop correction
for pure gauge theory, due to a gauge anomaly arising from the all-plus
helicity amplitude on twistor space.
In order to remedy this, a fourth-order scalar field
that couples to the Yang-Mills topological term was introduced
in refs.~\cite{Costello:2021bah, Costello:2022upu,Costello:2022wso}.
However, the mechanism can only cancel double-trace contributions, and
so it is necessary for the gauge group to not have an independent
quartic Casimir structure.  Alternatively, the anomaly can be cured by
introducing fermions in special representations of the gauge
group~\cite{Costello:2023vyy}.  In particular, the requirement is that
the quartic Casimir in the adjoint representation is exactly that in the
(real) representation $R$
\be
\tr_G(X^4) = \tr_R(X^4).
\ee
For $SU(N)$ guage theory, one such example of this type of representation is
$R_0$ given in eq.~(\ref{therep}).

With this choice of matter representation, the one-loop OPEs are
associative. Therefore, the chiral algebra exists for this theory and
can be used to compute form factors. In fact, associativity constrains
all form factors of self-dual Yang-Mills (plus matter) to be rational
functions, with poles only in the spinor products $\spa{ij}$.
The chiral algebra OPEs determine all possible poles in the form factor,
and the residues of these poles are chiral algebra correlators that have
fewer external states or are at lower loop order. In this way, one can
determine the $n$-point form factors inductively.
	
The form factor of most interest is the one with the operator
\be
 \frac{1}{2}\tr(B \wedge B),
\ee
inserted at the origin,\footnote{
We mean the origin in position space $x$.  The $x$-dependence
of the correlator is $\propto \exp(i\sum_{j=1}^n k_j\cdot x)$ where
$k_j$ are the gluon momenta.}
where $B$ is the adjoint-valued anti-self-dual
two-form appearing in the sdYM Lagrangian~\cite{Chalmers:1996rq},
\be
{\cal L}_{\rm sdYM} = \tr( B \wedge F ).
\ee
Deforming the self-dual Lagrangian by $\tfrac{1}{2}g^2\tr(B\wedge B)$
and integrating out $B$ yields the regular Yang-Mills Lagrangian,
up to a topological term which does not affect the perturbation theory.
So form factors of self-dual Yang-Mills with the operator
$\tfrac{1}{2}\tr(B\wedge B)$ inserted at the origin are amplitudes of
ordinary Yang-Mills theory.
	
Using the CCA bootstrap, massless QCD amplitudes with matter in the
representation \eqref{therep} were computed at
tree level~\cite{Costello:2022wso}, one loop~\cite{Costello:2023vyy, Costello:2022upu},
and two loops~\cite{Costello:2023vyy} for the two-minus, one-minus,
and all-plus helicity configurations, respectively.
The two-loop all-plus four-point sdYM form factor is\footnote[1]{
  The double-trace term was not provided in ref.~\cite{Costello:2023vyy}.
  However, Appendix B of ref.~\cite{Costello:2023vyy} outlines the
  computation of the color factors, so that one can keep track of
  the double-trace terms if desired relatively easily;
  see Appendix~\ref{sec:dtcomputation}.}
\be
\label{KC4pt}
\mathcal{A}_{4,\text{sdYM}}^\text{2-loop}
= g^6 \Bigl[ A^\text{2-loop}_{4;1,\text{sdYM }} \big( \tr(1234) + \tr(1432) \big)
    + A^\text{2-loop}_{4;3,\text{sdYM}}\tr(12)\tr(34) \Bigr]
+ \mathcal{C}(234),
\ee
where
\begin{align}
\begin{split}
A^\text{2-loop}_{4;1,\text{sdYM}}
=&~\frac{i}{(4\pi)^4} \Biggl[ (6N-4-8N^{-1})
    \bigg( \frac{\spb{12}\spb{34}}{\spa{12}\spa{34}}
          +\frac{\spb{14}\spb{23}}{\spa{14}\spa{23}} \bigg)
-(4+8N^{-1})\frac{\spb{13}\spb{24}}{\spa{13}\spa{24}}
	\\
& - 2 \frac{\spb{12}\spb{34}}{\spa{12}\spa{34}}
       \frac{\spa{13}\spa{24}+\spa{14}\spa{23}}{\spa{12}\spa{34}}
- 2 \frac{\spb{14}\spb{23}}{\spa{14}\spa{23}}
    \frac{\spa{13}\spa{24}+\spa{12}\spa{34}}{\spa{14}\spa{23}}
      \Biggr]          
\end{split}
\label{KC_A41}
\end{align}
and
\begin{align}
A_{4;3,\text{sdYM}}^\text{2-loop} =
\frac{8i}{(4\pi)^4} (1+N^{-1})
		\bigg(
		  \frac{\spb{12}\spb{34}}{\spa{12}\spa{34}}
		+ \frac{\spb{13}\spb{24}}{\spa{13}\spa{24}}
		+ \frac{\spb{14}\spb{23}}{\spa{14}\spa{23}}
		\bigg) \,.
\label{KC_A43}
\end{align}
This expression can be simplified using the Schouten spinor identity
and four-point momentum conservation,
which includes the result that $\rho$ is totally symmetric,\footnote{%
Our overall normalization of form factors and amplitudes differs
from ref.~\cite{Costello:2023vyy} by a factor of $i$.}
\be
\frac{\rho}{i} = \frac{\spb{12}\spb{34}}{\spa{12}\spa{34}}
        	= \frac{\spb{13}\spb{24}}{\spa{13}\spa{24}}
		= \frac{\spb{14}\spb{23}}{\spa{14}\spa{23}} \,.
\label{rhosym}
\ee
Then eqs.~\eqref{KC_A41} and \eqref{KC_A43}
collapse to eq.~\eqref{KCformula}.
However, when computing $n$-point form factors based on lower-point ones,
one must remember \emph{not} to use lower-point momentum conservation to
simplify the lower-point form factors, as it is the sum of the $n$
gluon momenta that is conserved, not a subset of them.

With this in mind, the $n$-point color-ordered amplitude is constructed
recursively, based on eqs.~\eqref{KC_A41} and \eqref{KC_A43},
and is given by
\begin{align}
\label{KCnpt}
\begin{split}
\mathcal{A}^\text{2-loop}_{n,\text{sdYM}} &= g^{n+2} \Biggl[
  \sum_{\sigma\in S_n/\mathbb{Z}_n}\tr(\sigma_1\cdots\sigma_n)
\\
   & \hskip1.5cm \times \sum_{1\leq i<j<k<l\leq n}
       A^\text{2-loop}_{4;1,\text{sdYM}}(\sigma_i,\sigma_j,\sigma_k,\sigma_l)
    \frac{\spa{\sigma_i\sigma_j}\spa{\sigma_j\sigma_k}
          \spa{\sigma_k\sigma_l}\spa{\sigma_l\sigma_i}}
          {\spa{\sigma_1\sigma_2}\spa{\sigma_2\sigma_3}
          \cdots\spa{\sigma_n\sigma_1}}
\\
&+\sum_{c=3}^{\lfloor n/2\rfloor + 1}\sum_{\sigma\in S_n/S_{n;c}}
\tr(\sigma_1\cdots\sigma_{c-1})
\tr(\sigma_c\cdots\sigma_n)
\sum_{1\leq i<j<k<l\leq n}
A_{n;c,\text{sdYM}}^\text{2-loop}(\sigma_i,\sigma_j,\sigma_k,\sigma_l) \Biggr]\,,
\end{split}
\end{align}
where $S_{n;c}$ is the subgroup of $S_n$ consisting of permutations that
keep the double-trace structure $\tr(1,\dotsc,c-1)\tr(c,\dotsc,n)$ invariant.
$A_{n,c,\text{sdYM}}^\text{2-loop}(i,j,k,l)$ is the kinematic factor
that multiplies this double-trace structure for the form factor with
energy-level-1 insertions at $i,j,k,l$
(as explained in Appendix \ref{sec:dtcomputation}).
It is defined as
\be
\label{dtterm}
A_{n;c,\text{sdYM}}^\text{2-loop}(i,j,k,l)=
\frac{	A_{4;3,\text{sdYM}}^\text{2-loop}(i,j,k,l)\spa{ij}^2\spa{kl}^2}
		{\spa{12}\spa{23}\cdots\spa{c-1,1}
			\spa{c,c+1}\spa{c+1,c+2}\cdots\spa{n,c}}
\ee
for $1\leq i<j\leq c-1$ and $c\leq k<l\leq n$, and it is zero otherwise.
In Appendix \ref{sec:dtcomputation}, we prove eq.~\eqref{dtterm} using the
CCA bootstrap. 

Note that for fermionic matter in $R_0$ there is no triple trace contribution,
which would be present generically.  The triple-trace cancellation is
a consequence of the recursive construction, and its absence for $n=4$
since $\tr(t^a)=0$ in $SU(N)$.

We wish to check
eqs.~(\ref{KC_A41})--(\ref{dtterm}) in the
simplest case, $n=4$, via an alternative method.
We will use the fact that the two-loop four-gluon amplitudes
were computed in QCD in dimensional
regularization~\cite{Bern:2000dn,Bern:2002tk} in a color-decomposed
form which makes it straightforward to modify the fermion representation
to $R_0$.

Before doing the two-loop color algebra, we first warm up by computing
the one-loop all-plus $n$-point amplitude, which vanishes (non-trivially)
in this theory due to the anomaly cancellation~(\ref{eq:cancelanom}).

\section{The One-loop Amplitude}
\label{sec:oneloop}

Here, we compute the one-loop all-plus $n$-point amplitude for massless QCD
with matter in the representation~\eqref{therep}. Color-decomposition
plays a crucial role in this computation. We begin by reviewing the
color-decomposition of one-loop $n$-gluon amplitudes in QCD for gauge
group $SU(N)$ with matter in the representation $N_F ( F\oplus\bar{F} )$,
where $N_F$ is the number of quark flavors.

\subsection{One-loop in QCD}
\label{sec:oneloopQCD}

The one-loop $n$-gluon QCD amplitude can be color-decomposed
as~\cite{Bern:1990ux}
\begin{align}
\label{1loopdecomp}
\begin{split}
\mathcal{A}_{n,\text{QCD}}^\text{1-loop} = g^n \Bigg[
&N\sum_{\sigma\in S_n/\mathbb{Z}_n}\tr(\sigma_1\cdots\sigma_n)
A_n^{[1]}(\sigma_1,\dotsc,\sigma_n)
\\
&+\sum_{c=3}^{\lfloor n/2 \rfloor+1}\sum_{\sigma\in S_n/S_{n;c}}
 \tr(\sigma_1\cdots\sigma_{c-1})\tr(\sigma_c\cdots\sigma_n)
A_{n;c}(\sigma_1,\dotsc,\sigma_n)
\\
&+ N_F \sum_{\sigma\in S_n/\mathbb{Z}_n}\tr(\sigma_1\cdots\sigma_n)
A_n^{[1/2]}(\sigma_1,\dotsc,\sigma_n)
		\Bigg],
\end{split}
\end{align}
where the $A_{n;c}$ are the subamplitudes.
The superscript $[j]$ denotes the spin of the particle circulating
in the loop, $j=1/2$ or $1$.  The subamplitudes $A_n^{[j]}$ are color-ordered.

The subleading subamplitudes $A_{n;c}$ are obtained from the leading ones
$A_n^{[1]}$ through the permutation sum~\cite{Bern:1994zx,DelDuca:1999rs}
\be
A_{n;c}(\alpha,\beta) = (-1)^{|\beta|}
\sum_{\sigma\,\in\,\alpha\shuffle\beta^T}A_n^{[1]}(\sigma_1,\ldots,\sigma_n),
\label{Ancformula}
\ee
where $\alpha=(1,2,\ldots,c-1)$ and $\beta=(c,c+1,\ldots, n)$
are cyclicly ordered lists, and $\beta^T=(n,\ldots,c+1,c)$
is the reverse ordering, with the understanding that $\alpha$ and $\beta^T$
are actually equivalence classes under cyclic permutations of
their arguments, i.e.
\begin{align}
\alpha &= \{(1,2,\dotsc, c-1),(2,\dotsc,c-1,1),\dotsc,(c-1,1,\dotsc,c-2)\},
 \\
\beta^T &= \{(n,n-1,\dotsc,c),(n-1,\dotsc,c,n),\dotsc,(c,n,\dotsc,c+1)\}.
\end{align}
The symbol $\alpha\shuffle\beta^T$ denotes the cyclic shuffle product,
which is the set of all permutations up to cycles of $\{1,2,\ldots,n\}$
that preserve the cyclic ordering of $\alpha$ and $\beta^T$,
while allowing all possible relative orderings of the elements of $\alpha$
with respect to the elements of $\beta^T$. For example,
letting $\alpha=(1,2,3)$ and $\beta=(4,5)$, we have
\begin{align}
 \alpha\shuffle\beta^T = \{
 &(1,2,3,4,5), (1,2,4,3,5), (1,4,2,3,5),
             (1,2,4,5,3), (1,4,2,5,3), (1,4,5,2,3), \nonumber\\
 &(1,2,3,5,4), (1,2,5,3,4),(1,5,2,3,4),
             (1,2,5,4,3), (1,5,2,4,3), (1,5,4,2,3) \}.
\end{align}
Again, it is understood that the lists within this set are equivalence
classes under cyclic permutations of their arguments.
	
Another color decomposition also exists for the
gluon (adjoint) contribution, in terms of traces over generators in
the adjoint representation of $SU(N)$~\cite{DelDuca:1999rs}
\be
\label{ringdecomp}
\mathcal{A}_{n,\text{QCD}}^\text{1-loop}
= \frac{g^n}{2} \sum_{\sigma\in S_n/\mathbb{Z}_n} \Bigl[
  \tr_G(\sigma_1\ldots\sigma_n) A_n^{[1]}(\sigma_1,\ldots,\sigma_n)
  + 2 N_F\tr(\sigma_1\ldots\sigma_n)A_n^{[1/2]}(\sigma_1,\ldots,\sigma_n)
  \Bigr] \,.
\ee
The factor of $1/2$ accounts for a reflection identity
$\tr_G(i_ni_{n-1}\cdots i_1) = (-1)^n \tr_G(i_1i_2\cdots i_n)$,
which implies a reflection identity on the color-ordered subamplitude
$A_n^{[1]}$ (which also holds for $A_n^{[1/2]}$):
\be
 A_n^{[j]}(n,n-1,\ldots,1) = (-1)^n \, A_n^{[j]}(1,2,\ldots,n).
\label{reflid}
\ee
The sum in eq.~\eqref{ringdecomp} includes $\sigma^T$
for all $\sigma\in S_n/\mathbb{Z}_n$, so the factor of $1/2$ is needed.
	
The equivalence of eqs.~\eqref{1loopdecomp} and~\eqref{Ancformula}
with eq.~\eqref{ringdecomp} can be
seen by representing the adjoint representation $G$ in terms of
fundamental representations, $G \oplus 1 \cong F\otimes\bar{F}$.
Evaluating the $F\otimes\bar{F}$ traces we have,
\begin{align}
\tr_G(1\cdots n) = \tr_{F\otimes\bar{F}}(1\cdots n)
 =&\sum_{I\subset(1,\cdots,n)}\tr(I)\tr_{\bar{F}}(I^c)
\nonumber\\
=&~N \tr(1\ldots n) + (-1)^n N \tr(n\ldots 1)
\nonumber\\
&+\sum_{\emptyset\neq I\subsetneq(1,\ldots,n)} (-1)^{|I^c|}
         \tr(I)\tr((I^c)^T),
\label{adjointdecomp}
\end{align}
where $I^c$ is the complement of the sublist $I$.
The notation $I\subset(1,\dotsc,n)$ means that $I$ is a sublist of
$(1,\dotsc,n)$ with respect to which $I$ is ordered.
(In $SU(N)$, $\tr(t^a)=0$, so one can drop the cases with $|I|=1$ and
$|I|=n-1$.)
This relation has a nice diagrammatic representation in terms of color
graphs using the double-line notation, as shown in fig.~\ref{fig:g=ffbar}.
As a reminder, in the double-line notation the rule is to sum all $2^n$
ways of attaching the $n$ external lines to either the inner or
outer ring of the annulus, with a minus sign for each attachment to
the inner ($\bar{F}$) ring.
	
	\begin{figure}[h]
	\centering
	\be
	\nonumber
	\begin{tikzpicture}[baseline={(0,0)},scale=0.75]
		\begin{feynman}
			\vertex (g1) at (0,-2);
			\vertex (g2) at (0,2);
			\vertex (g3) at (4,2);
			\vertex (g4) at (4,-2);
			\vertex (v1) at (1,-1);
			\vertex (v2) at (1,1);
			\vertex (v3) at (3,1);
			\vertex (v4) at (3,-1);
			\diagram*[edges=gluon]{
				(g1) -- (v1),
				(v2) -- (g2),
				(g3) -- (v3),
				(v4) -- (g4),
			(v1) -- [quarter right] (v4) -- [quarter right] (v3) 
			     -- [quarter right] (v2) -- [quarter right] (v1),
			};
		\end{feynman}
	\end{tikzpicture}
		=
	\begin{tikzpicture}[baseline={(0,0)},scale=0.75]
		\begin{feynman}
			\vertex (g1) at (0,-2);
			\vertex (g2) at (0,2);
			\vertex (g3) at (4,2);
			\vertex (g4) at (4,-2);
			\vertex (v1) at (1,-1);
			\vertex (v2) at (1,1);
			\vertex (v3) at (3,1);
			\vertex (v4) at (3,-1);
			\vertex (w1) at (1.25,-0.75);
			\vertex (w2) at (1.25,0.75);
			\vertex (w3) at (2.75,0.75);
			\vertex (w4) at (2.75,-0.75);
			\diagram*{
				(g1) -- [gluon] (v1),
				(v2) -- [gluon] (g2),
				(g3) -- [gluon] (v3),
				(v4) -- [gluon] (g4),
				(v1) -- [quarter left, fermion] (v2) 
					 -- [quarter left, fermion] (v3) 
				     -- [quarter left, fermion] (v4) 
				     -- [quarter left, fermion] (v1),
				(w1) -- [quarter left, anti fermion] (w2) 
					 -- [quarter left, anti fermion] (w3) 
				     -- [quarter left, anti fermion] (w4) 
				     -- [quarter left, anti fermion] (w1),
			};
		\end{feynman}
\end{tikzpicture}
\ee
\caption{Graphical representation of the $SU(N)$ identity
$G\oplus1\cong F\otimes\bar{F}$. The diagram on the right is evaluated
by summing over all $2^n$ ways of attaching $n$ external legs to
either ring of the annulus, with a minus sign for each attachment
to the inner ($\bar{F}$) ring.}
\label{fig:g=ffbar}
\end{figure}
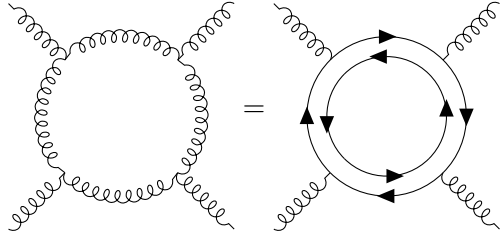

When all external gluons have positive helicities, the color-ordered
subamplitudes are finite, rational functions of spinor products $\spa{ij}$
and $\spb{ij}$ given by~\cite{Mahlon:1993si,Bern:1993qk}
\begin{align}
A_n^{[1]}(1,2,\dotsc,n) &= -\frac{i}{48\pi^2}
\sum_{1\leq i_1<i_2<i_3<i_4\leq n}
\frac{\spa{i_1i_2}\spb{i_2i_3}\spa{i_3i_4}\spb{i_4i_1}}
{\spa{12}\spa{23}\cdots\spa{n1}} \,,
\label{1looppartials} \\
\label{1loopsusy}
A_n^{[1/2]}(1,2,\dotsc,n)&=-A_n^{[1]}(1,2,\dotsc,n),
\end{align}
where we have taken $N_p=2$ for $A_n^{[1]}$,
where $N_p$ is the number of bosonic states minus fermionic states.
Eq.~(\ref{1loopsusy}) is a supersymmetry
Ward identity (SWI)~\cite{Grisaru:1976vm,Grisaru:1977px,Parke:1985pn}
which holds in $D=4$.
At two loops, we will need to use dimensional regularization in
$D=4-2\epsilon$ spacetime dimensions, and we will need the one-loop
result for $n=4$ to higher orders in $\epsilon$.  For this purpose,
a formula for the subamplitudes in terms of a dimensionally-regulated
box integral is given in section~\ref{sec:IRsubtraction}.

\subsection{Including Matter in $8F\oplus8\bar{F}\oplus\asym\oplus\asyma$}
\label{sec:including matter in R}
	
According to ref.~\cite{Costello:2023vyy}, including matter in the
representation \eqref{therep} should nullify the one-loop all-plus amplitude.
This vanishing implies linear relations among the subamplitudes,
which we wish to elucidate. To do so, we need to compute traces over
the antisymmetric tensor representation in terms of traces over
fundamental representation generators.
	
For this computation, we can simply replace the fermion loops in the
fundamental representation that appear in the one-loop color graphs
with loops in the representation~\eqref{therep}. This replacement
is permitted for the following reason. Every Feynman diagram can be
written as the product of a color factor and a kinematic factor.
The Jacobi identity on the color factors can be used to remove color
graphs with nontrivial trees attached to the loop~\cite{DelDuca:1999rs},
and thereby rewrite the matter contribution as a sum of permutations of
the ``ring'' color diagram in fig.~\ref{fig:fermion1loop}.
Because the Jacobi identity is independent of the choice of representation
of the fermion loop, we arrive at the same sum over color diagrams,
with the same choice of fermion representation with which we began,
without affecting the final kinematic factors.  That is to say,
$A_n^{[j]}$ depends solely on the spin of the particle propagating in the
loop, not the representation of the Lie algebra in which it resides.

	\begin{figure}[h]
	\centering
	\begin{tikzpicture}[scale=0.75]
		\begin{feynman}
				\vertex (g1) at (0,-2);
				\vertex (g2) at (0,2);
				\vertex (g3) at (4,2);
				\vertex (g4) at (4,-2);
				\vertex (v1) at (1,-1);
				\vertex (v2) at (1,1);
				\vertex (v3) at (3,1);
				\vertex (v4) at (3,-1);
				\diagram*{
					(g1) -- [gluon] (v1),
					(v2) -- [gluon] (g2),
					(g3) -- [gluon] (v3),
					(v4) -- [gluon] (g4),
					(v1) -- [quarter left, fermion] (v2) 
			-- [quarter left, fermion, edge label=\(R\)] (v3) 
					-- [quarter left, fermion] (v4) 
					-- [quarter left, fermion] (v1),
				};
		\end{feynman}
	\end{tikzpicture}
\caption{The one-loop color diagram for matter in an arbitrary
          representation $R$ of $SU(N)$.}
\label{fig:fermion1loop}
\end{figure}
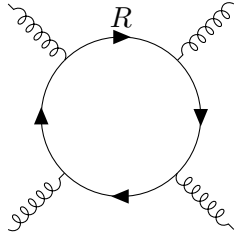
	
In other words, the contribution from matter in the
representation~\eqref{therep} to the one-loop amplitude is
\be
g^n\sum_{\sigma\in S_n/\mathbb{Z}_n}
\tr_{R_0}(\sigma_1\cdots\sigma_n) \, A_n^{[1/2]}(\sigma_1,\ldots,\sigma_n).
\ee
The color diagram $\tr_{R_0}(\sigma_1\cdots\sigma_n)$
associated to $A_n^{[1/2]}$ for this specific choice
of representation is shown in fig.~\ref{fig:R1loop}.
The rectangle covering the lines appearing in the diagrams
denotes antisymmetrization of those lines, as depicted in
fig.~\ref{fig:asym}.  The trace over $R_0$ in terms of traces
over the fundamental is worked out in Appendix \ref{sec:colorid}, and is
\begin{align}
 \begin{split}
\tr_{R_0}(t^{a_1}\cdots t^{a_n})
=&~ 8\tr(1\cdots n) + 8\tr_{\bar{F}}(1\cdots n)
   + \tr_{\asym}(1\cdots n) + \tr_{\asyma}(1\cdots n)
\\
=&~ 8\tr(1\cdots n) + (-1)^n8\tr(n\cdots 1)
  +N\tr(1\cdots n) + (-1)^nN\tr(n\cdots 1)
\\
&-\frac{1}{2}\sum_{I\subset(1,\dotsc,n)}
\big[ \tr(I\cdot I^c) + (-1)^n\tr((I\cdot I^c)^T) \big]
\\
&+\frac{1}{2}\sum_{\emptyset\neq I\subsetneq(1,\dotsc,n)}
\big[ \tr(I)\tr(I^c) + (-1)^n\tr(I^T)\tr((I^c)^T) \big],
\end{split}
\label{R0decomp}
\end{align}
where $I\cdot I^c$ means to concatenate the lists $I$ and $I^c$.

\begin{figure}[h]
		\centering
		\be
		\nonumber
		8
		\begin{tikzpicture}[baseline={(0,0)},scale=0.75]
			\begin{feynman}
				\vertex (g1) at (0,-2);
				\vertex (g2) at (0,2);
				\vertex (g3) at (4,2);
				\vertex (g4) at (4,-2);
				\vertex (v1) at (1,-1);
				\vertex (v2) at (1,1);
				\vertex (v3) at (3,1);
				\vertex (v4) at (3,-1);
				\diagram*{
					(g1) -- [gluon] (v1),
					(v2) -- [gluon] (g2),
					(g3) -- [gluon] (v3),
					(v4) -- [gluon] (g4),
					(v1) -- [quarter left, fermion] (v2) 
					-- [quarter left, fermion] (v3) 
					-- [quarter left, fermion] (v4) 
					-- [quarter left, fermion] (v1),
				};
			\end{feynman}
		\end{tikzpicture}
		+8
		\begin{tikzpicture}[baseline={(0,0)},scale=0.75]
			\begin{feynman}
				\vertex (g1) at (0,-2);
				\vertex (g2) at (0,2);
				\vertex (g3) at (4,2);
				\vertex (g4) at (4,-2);
				\vertex (v1) at (1,-1);
				\vertex (v2) at (1,1);
				\vertex (v3) at (3,1);
				\vertex (v4) at (3,-1);
				\diagram*{
					(g1) -- [gluon] (v1),
					(v2) -- [gluon] (g2),
					(g3) -- [gluon] (v3),
					(v4) -- [gluon] (g4),
					(v1) -- [quarter right, fermion] (v4) 
					-- [quarter right, fermion] (v3) 
					-- [quarter right, fermion] (v2) 
					-- [quarter right, fermion] (v1),
				};
			\end{feynman}
		\end{tikzpicture}
		+
		\begin{tikzpicture}[baseline={(0,0)},scale=0.75]
		\begin{feynman}
			\vertex (g1) at (0,-2);
			\vertex (g2) at (0,2);
			\vertex (g3) at (4,2);
			\vertex (g4) at (4,-2);
			\vertex (v1) at (1,-1);
			\vertex (v2) at (1,1);
			\vertex (v3) at (3,1);
			\vertex (v4) at (3,-1);
			\vertex (w1) at (1.25,-0.75);
			\vertex (w2) at (1.25,0.75);
			\vertex (w3) at (2.75,0.75);
			\vertex (w4) at (2.75,-0.75);
			\vertex (b1) at (1.65,-1.75);
			\vertex (b2) at (1.65,-0.75);
			\vertex (b3) at (2.35,-0.75);
			\vertex (b4) at (2.35,-1.75);
			\diagram*{
				(g1) -- [gluon] (v1),
				(v2) -- [gluon] (g2),
				(g3) -- [gluon] (v3),
				(v4) -- [gluon] (g4),
				(v1) -- [quarter left, fermion] (v2) 
				     -- [quarter left, fermion] (v3) 
				     -- [quarter left, fermion] (v4) 
				     -- [quarter left, fermion] (v1),
				(w1) -- [quarter left, fermion] (w2) 
				     -- [quarter left, fermion] (w3) 
				     -- [quarter left, fermion] (w4) 
				     -- [quarter left, fermion] (w1),
				(b1) -- (b2) -- (b3) -- (b4) -- (b1),
			};
		\end{feynman}
		\end{tikzpicture}
		+
		\begin{tikzpicture}[baseline={(0,0)},scale=0.75]
			\begin{feynman}
				\vertex (g1) at (0,-2);
				\vertex (g2) at (0,2);
				\vertex (g3) at (4,2);
				\vertex (g4) at (4,-2);
				\vertex (v1) at (1,-1);
				\vertex (v2) at (1,1);
				\vertex (v3) at (3,1);
				\vertex (v4) at (3,-1);
				\vertex (w1) at (1.25,-0.75);
				\vertex (w2) at (1.25,0.75);
				\vertex (w3) at (2.75,0.75);
				\vertex (w4) at (2.75,-0.75);
				\vertex (b1) at (1.65,-1.75);
				\vertex (b2) at (1.65,-0.75);
				\vertex (b3) at (2.35,-0.75);
				\vertex (b4) at (2.35,-1.75);
				\diagram*{
					(g1) -- [gluon] (v1),
					(v2) -- [gluon] (g2),
					(g3) -- [gluon] (v3),
					(v4) -- [gluon] (g4),
					(v1) -- [quarter right, fermion] (v4) 
					     -- [quarter right, fermion] (v3) 
					     -- [quarter right, fermion] (v2) 
					     -- [quarter right, fermion] (v1),
					(w1) -- [quarter right, fermion] (w4) 
					     -- [quarter right, fermion] (w3) 
					     -- [quarter right, fermion] (w2) 
					     -- [quarter right, fermion] (w1),
					(b1) -- (b2) -- (b3) -- (b4) -- (b1),
				};
			\end{feynman}
		\end{tikzpicture}
\ee
\caption{The one-loop color diagram for matter in the representation
  $R_0=8F\oplus8\bar{F}\oplus\asym\oplus\asyma$.}
\label{fig:R1loop}
\end{figure}
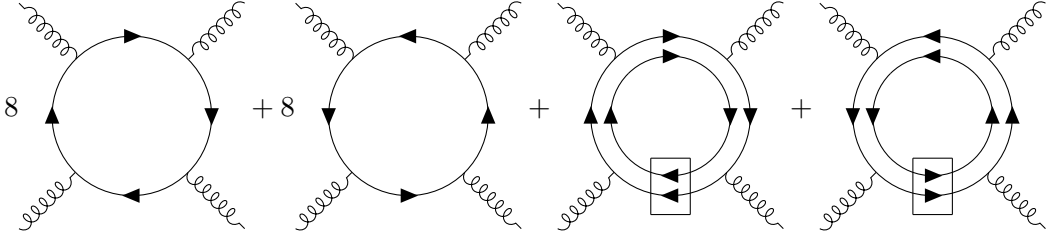

 	\begin{figure}
		\centering
		\be
			\nonumber
			\begin{tikzpicture}[baseline={(0,0)},scale=1]
				\begin{feynman}
					\vertex (v1) at (1,-0.2);
					\vertex (v2) at (3,-0.2);
					\vertex (w1) at (1,0.5);
					\vertex (w2) at (3,0.5);
					\vertex (b1) at (1.65,-0.4);
					\vertex (b2) at (1.65,0.7);
					\vertex (b3) at (2.35,0.7);
					\vertex (b4) at (2.35,-0.4);
					\diagram*{
						(v1) -- [fermion] (v2),
						(w1) -- [fermion] (w2),
					(b1) -- (b2) -- (b3) -- (b4) -- (b1),
					};
				\end{feynman}
			\end{tikzpicture}
			=
			\frac{1}{2}
			\Bigg(
			\begin{tikzpicture}[baseline={(0,0)},scale=1]
				\begin{feynman}
					\vertex (v1) at (1,-0.2);
					\vertex (v2) at (3,-0.2);
					\vertex (w1) at (1,0.5);
					\vertex (w2) at (3,0.5);
					\diagram*{
						(v1) -- [fermion] (v2),
						(w1) -- [fermion] (w2),
					};
				\end{feynman}
			\end{tikzpicture}
			-
			\begin{tikzpicture}[baseline={(0,0)},scale=1]
				\begin{feynman}
					\vertex (v1) at (1,-0.2);
					\vertex (v2) at (3,-0.2);
					\vertex (w1) at (1,0.5);
					\vertex (w2) at (3,0.5);
					\vertex (v1p) at (1.75,0.0875);
					\vertex (w2p) at (2.25,0.2125);
					\diagram*{
						(v1) -- [fermion] (v1p),
						(w2p) -- [fermion] (w2),
						(w1) -- [fermion] (v2),
					};
				\end{feynman}
			\end{tikzpicture}
			\Bigg)
		\ee
		\caption{Graphical representation of the antisymmetric tensor
                product of the fundamental representation in terms of
                two fundamental lines.}
		\label{fig:asym}
	\end{figure}

Combining the decomposition~(\ref{adjointdecomp}) of the adjoint
pure-gluon contribution with the $R_0$ matter contribution~(\ref{R0decomp})
yields
\begin{align}
\label{trGR}
\begin{split}
  \tr_G(1\cdots n)&A_n^{[1]}(1,\dotsc,n)
  + \tr_{R_0}(1\cdots n)A_n^{[1/2]}(1,\dotsc,n)
\\
=&~ -8\tr(1\cdots n)A_n^{[1]}(1,\dotsc,n)
    - 8\tr(n\cdots 1) A_n^{[1]}(n,\dotsc,1)
\\
&+\frac{1}{2}\sum_{I\subset(1,\dotsc,n)}
 \bigl[ \tr(I\cdot I^c)A_n^{[1]}(1,\dotsc,n) 
 + \tr((I\cdot I^c)^T)A_n^{[1]}(n,\dotsc,1) \bigr]
\\
&+\frac{1}{2}\sum_{\emptyset\neq I\subsetneq(1,\dotsc,n)}
		\big[
		2\tr(I)\tr((I^c)^T)-\tr(I)\tr(I^c)-\tr(I^T)\tr((I^c)^T)
		\big]A_n^{[1]}(1,\dotsc,n),
\end{split}
\end{align}
where we have used the SWI~\eqref{1loopsusy}
and the reflection identity~\eqref{reflid} obeyed by the subamplitude.
The full amplitude is then given by the sum over all permutations
on $n$ letters, modulo permutations related by cycles and reflections. 
	
We define the subamplitude $A_n^{R_0}(1,\dotsc,n)$ to be the kinematic
factor multiplying the single-trace color factor $\tr(1,\dotsc,n)$
in eq.~\eqref{trGR}. It is given by
\be
\label{AGR}
A_n^{R_0}(1,\dotsc,n) = - 8 A_n^{[1]}(1,\dotsc,n)
  + \sum_{k=1}^n\sum_{\sigma\,\in\,\alpha_k\shuffle\beta_k}A_n^{[1]}(1,\sigma),
\ee
where $\alpha_k=(2,\dotsc,k)$ and $\beta_k=(k+1,\dotsc,n)$.
The first term comes from the trace over eight copies of the fundamental.
The remaining terms come from the exchange term in the trace over the
antisymmetric tensor representation,
\be
\label{trproj}
\frac{1}{2}\tr_{F\otimes F}(1\cdots nP)
 = \frac{1}{2}\sum_{I\subset(1,\dotsc,n)}\tr(I\cdot I^c),
\ee
where $P$ is the permutation operator that exchanges the two $F$
representations.
In particular, the sum over $k$ appears since the list
$(1,\dotsc,k)=(1,\alpha_k)$ appears in the sum in eq.~\eqref{trproj}
for all $1\leq k\leq n$.  In Appendix \ref{sec:AGRproof}, we show
that the subamplitude $A_n^{R_0}(1,\dotsc,n)$ is given by eq.~\eqref{AGR}.

Since the full amplitude vanishes for the fermion representation $R_0$
and the traces over the generators are linearly independent in $SU(N)$
(up to dihedral symmetries), eq.~\eqref{AGR} must also vanish:
\be
\label{1loopreln}
0 = - 8 A_n^{[1]}(1,\dotsc,n)
    + \sum_{k=1}^n\sum_{\sigma\,\in\,\alpha_k\shuffle\beta_k}A_n^{[1]}(1,\sigma).
\ee
Remarkably, these relations are exactly the same all-plus relations
conjectured in ref.~\cite{Bjerrum-Bohr:2011jrh}.
Ref.~\cite{Bjerrum-Bohr:2011jrh} based their formula\footnote{
Note that the boundary terms
$k=1$ and $k=n$ have an empty set for $\alpha_k$ and for $\beta_k$,
respectively, so they each just give $A_n^{[1]}(1,\dotsc,n)$. Removing
them from the sum over $k$ puts the formula into the precise
form in ref.~\cite{Bjerrum-Bohr:2011jrh}.}
on a decomposition into kinematic diagrams containing a single totally symmetric
quartic vertex, and the remaining vertices are all cubic and
totally antisymmetric.  If one accepts that the twistor-space
anomaly cancellation implies the vanishing of the one-loop all-plus amplitude,
then one obtains a proof of this conjecture.
In a forthcoming paper~\cite{DMToAppear}, we analyze these all-plus relations,
relations among one-loop one-minus amplitudes, as well as their
connections to the twistor-space anomaly cancellation mechanism
that uses a fourth-order
pseudoscalar~\cite{Costello:2021bah,Costello:2022upu,Costello:2022wso}.

We can verify eq.~\eqref{1loopreln} for the case $n=4$.
The $n=4$ all-plus partial amplitude is
\be
A_4^{[1]}(1,2,3,4) = -\frac{i}{48\pi^2}
\frac{\spb{23}\spb{41}}{\spa{23}\spa{41}}  = - \frac{\rho}{48\pi^2} \,.
\label{allplus4}
\ee
This expression is totally symmetric, as shown in eq.~\eqref{rhosym}.
For general $n$, the number of terms appearing in the sum over $k$
in eq.~\eqref{1loopreln} is
\be
\sum_{k=1}^{n}\binom{n-1}{k-1}=2^{n-1},
\ee
counting multiplicities. So, for $n=4$, there are $2^3=8$ terms,
all of which are equal, thanks to the total symmetry of the four-point
subamplitude. These eight copies come with the opposite sign of the
$8$ terms not in the sum over $k$, resulting in a total of zero.

Because of the total kinematic symmetry of
eq.~\eqref{allplus4}, the above verification of eq.~\eqref{1loopreln}
for $n=4$ is equivalent to checking the anomaly cancellation
condition~\eqref{eq:cancelanom}; both involve the same symmetrized
trace over four generators in the appropriate representations.

For $n>4$, eq.~\eqref{1loopreln} is not so easily verified from
the explicit formula~\eqref{1looppartials}.
We have checked~\cite{DMToAppear}
that it holds for $n\leq 11$ by replacing all spinor
brackets with $3n-10$ independent momentum-invariants, using
a momentum-twistor-based
parametrization~\cite{Hodges:2009hk,Badger:2013gxa}.

So far we have discussed the consequence of the all-plus vanishing
for $R_0$ via the coefficient of the single trace $\tr(1\cdots n)$.
However, eq.~\eqref{trGR} also has a double-trace contribution,
which must vanish as well.  The relations among color-ordered amplitudes
that follow from this vanishing imply the vanishing of amplitudes with
three photons and $(n-3)$ gluons observed previously~\cite{Bern:1993qk}.
We will discuss these double-trace relations in a forthcoming
paper~\cite{DMToAppear}.


The vanishing of the one-loop amplitude in the $R_0$ theory suggests that
the two-loop amplitude should be finite and rational; indeed,
such behavior is found via the CCA bootstrap~\cite{Costello:2023vyy}.
However, we will see that eq.~\eqref{1loopreln} only holds at order $\eps^0$
in dimensional regularization; it fails at higher orders in $\eps$,
for the case $n=4$ (see section~\ref{sec:IRsubtraction}).
Consequently, the IR structure of the dimensionally-regulated
two-loop result is more intricate, and not even finite as $\eps\to0$.

\section{The 2-loop 4-gluon amplitude}
\label{sec:twoloopdimreg}

We now turn to the computation of the two-loop all-plus four-gluon
amplitude for fermions in the representation $R_0$.

Our starting point is the two-loop four-gluon amplitude in QCD, which is
given in ref.~\cite{Bern:2002zk} as 
\be
\mathcal{A}^\text{2-loop}_\text{4,QCD}
= \mathcal{A}_G^\text{adj} + \mathcal{A}^\text{fund}_F,
\ee
where $\mathcal{A}_G^\text{adj}$ is the adjoint gluon contribution
and $\mathcal{A}_F^\text{fund}$ is the fundamental matter contribution.
Each particle contribution above can be decomposed into a sum of
``parent'' diagrams,
\be
\mathcal{A}^\text{rep}_X = g^6\sum_{D_i}
\Big[ (C_\text{rep})_{1234}^{D_i}A_{X1234}^{D_i}
    +(C_\text{rep})_{3421}^{D_i}A_{X3421}^{D_i} \Big] + \mathcal{C}(234),
\ee
where each $D_i$ corresponds to a parent diagram.
The subscript $X\in\{G,F\}$ denotes either the pure-gluon contribution $G$,
or a fermion $F$ propagating in at least one of the two loops in the diagrams.
The quantities $C_\text{rep}$ denote the color factors associated to the
kinematic factors $A_X$, with ``$\text{rep}$'' signifying the gauge group
representation in which particle $X$ resides.
The $F$ parent diagrams
span the space of all independent four-gluon color-factors with a
fermion-loop contribution and a non-vanishing kinematic factor.
This result can be shown by applying the Jacobi identity suitably
to color diagrams containing triangle subdiagrams.
	
We want to compute the two-loop four-gluon amplitude with matter
in the representation $R_0$,
\be
\mathcal{A}_4^\text{2-loop} = \mathcal{A}^\text{adj}_G
                           + \mathcal{A}^{R_0}_F \,.
\ee
%

\subsection{Pure gauge contribution}
\label{sec:puregauge}

The color decomposition of the pure Yang-Mills two-loop four-point
amplitude is~\cite{Bern:2000dn,Bern:2002zk}
\be
\label{pureglu}
\mathcal{A}_G^\text{adj}=g^6\Big(C_{1234}^PA_{G1234}^P+C_{3421}^PA_{G3421}^P
+ C_{1234}^{NP}A_{G1234}^{NP}+C_{3421}^{NP}A_{G3421}^{NP} \Big) + \mathcal{C}(234),
\ee
where $C^P_{1234}$ and $C^{NP}_{1234}$ are color factors given by the
planar and non-planar parent diagrams in fig.~\ref{fig:2loopG}.
They are computed by dressing each vertex and each propagator with
the diagrammatic rules given in eq.~\eqref{diagramrules}.
	
	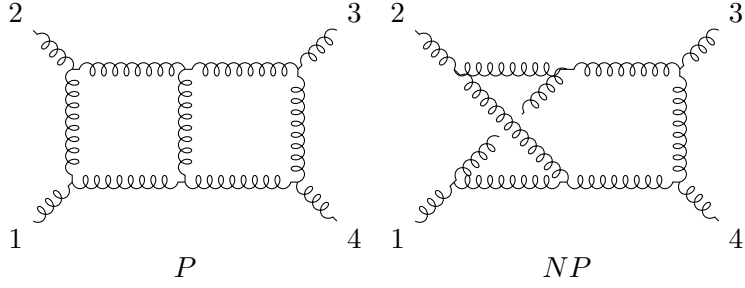
\begin{figure}
	\centering
	\begin{tikzpicture}[scale=0.75]
		\begin{feynman}
			\vertex (g1) at (-3,-2) {\(1\)};
			\vertex (g2) at (-3,2) {\(2\)};
			\vertex (g3) at (3,2) {\(3\)};
			\vertex (g4) at (3,-2) {\(4\)};
			\vertex (v1) at (-2,-1);
			\vertex (v2) at (-2,1);
			\vertex (v3) at (2,1);
			\vertex (v4) at (2,-1);
			\vertex (v5) at (0,1);
			\vertex (v6) at (0,-1);
			\vertex (g6) at (0,-2.5) {\(P\)};
			\diagram*[edges=gluon]{
				(g1) -- (v1),
				(v2) -- (g2),
				(g3) -- (v3),
				(v4) -- (g4),
				(v1) -- (v6)
				     -- (v4) 
				     -- (v3) 
				     -- (v5)
				     -- (v2) 
				     -- (v1),
				(v5) -- (v6)
				     -- [opacity=0] (g6),
			};
		\end{feynman}
	\end{tikzpicture}
	\begin{tikzpicture}[scale=0.75]
		\begin{feynman}
			\vertex (g1) at (-3,-2) {\(1\)};
			\vertex (g2) at (-3,2) {\(2\)};
			\vertex (g3) at (3,2) {\(3\)};
			\vertex (g4) at (3,-2) {\(4\)};
			\vertex (v1) at (-2,-1);
			\vertex (v2) at (-2,1);
			\vertex (v3) at (2,1);
			\vertex (v4) at (2,-1);
			\vertex (v5) at (0,1);
			\vertex (v6) at (0,-1);
			\vertex (g6) at (0,-2.5) {\(NP\)};
			\vertex (v7) at (-1.2,-0.2);
			\vertex (v8) at (-0.8,0.2);
			\diagram*[edges=gluon]{
				(g1) -- (v1),
				(v2) -- (g2),
				(g3) -- (v3),
				(v4) -- (g4),
				(v1) -- (v6)
				     -- (v2) 
				     -- (v5)
				     -- (v8),
				(v7) -- (v1),
				(v6) -- (v4)
				     -- (v3) 
				     -- (v5),
				(v6) -- [opacity=0] (g6),
			};
		\end{feynman}
	\end{tikzpicture}
	\caption{The planar ($P$) and non-planar ($NP$) parent
          color diagrams for the pure-gauge two-loop amplitude.}
	\label{fig:2loopG}
	\end{figure}

The color factors evaluate to
\begin{align}
\begin{split}
  C^P_{1234} =&~ (N^2+2)\big[\tr(1234) + \tr(1432)\big]
  + 2\big[\tr(1243)+\tr(1342)\big] 
\\
&- 4\big[\tr(1324)+\tr(1423)\big] +6N\tr(12)\tr(34) \,,
\end{split}
\\
\begin{split}
C^{NP}_{1234} =&~ 2\big[\tr(1234)+\tr(1432)+\tr(1243)+\tr(1342)\big]
			-4\big[\tr(1324)+\tr(1423)\big]
\\
&+2N\big[ 2 \, \tr(12)\tr(34) - \tr(13)\tr(24) - \tr(14)\tr(23) \big] \,.
\end{split}
\end{align}
These color factors have the following symmetries, which will prove
useful in section~\ref{sec:IRsubtraction}:
\begin{align}
\label{adjcolorsym}
\begin{split}
C^P_\text{1234} &= C^P_{3412} = C^P_{2143} = C^P_{4321} \,,
\\
C^{NP}_{1234} &= C^{NP}_{2134} = C^{NP}_{1243} = C^{NP}_{2143} \,.
\end{split}
\end{align}
	
The planar and non-planar primitive amplitudes are given by
\begin{align}
\label{gluprimamps}
\begin{split}
A^P_{G1234} =&~ \rho
\bigg\{		s(D_s-2)\mathcal{I}^P_4
\Big[ \lam_p^2\lam_{p+q}^2+\lam_q^2\lam_{p+q}^2 \Big] (s,t)
\\
&~~~~+\frac{(D_s-2)^2}{s}\mathcal{I}_4^\text{bow-tie}
  \Big[ \lam_p^2\lam_q^2 \big( (p+q)^2+s \big) \Big] (s,t)
		\bigg\} \,,
\\
A^{NP}_{G1234}=&~\rho s(D_s-2)\mathcal{I}^{NP}_4
\Big[ \lam_p^2\lam_q^2+\lam_p^2\lam_{p+q}^2+\lam_q^2\lam_{p+q}^2 \Big] \,,
\end{split}
\end{align}
where we have only included non-vanishing terms at ${\cal O}(\e^0)$
in the integral.
The three two-loop momentum integrals that appear above are the
planar double box integral $\mathcal{I}^P_4$, the non-planar double
box integral $\mathcal{I}^{NP}_4$, and the bow-tie integral
$\mathcal{I}^\text{bow-tie}_4$. They are shown in
fig.~\ref{fig:integrals} and are defined by,
\begin{multline}
 \mathcal{I}^P_4[\mathcal{P}(\lam_i,p,q,k_i)](s,t)
\\
= \int\frac{d^Dp}{(2\pi)^D}\frac{d^Dq}{(2\pi)^D}
\frac{\mathcal{P}(\lam_i,p,q,k_i)}
{p^2q^2(p+q)^2(p-k_1)^2(p-k_1-k_2)^2(q-k_4)^2(q-k_3-k_4)^2},
\end{multline}
\begin{multline}
\mathcal{I}^{NP}_4[\mathcal{P}(\lam_i,p,q,k_i)](s,t)
\\
= \int\frac{d^Dp}{(2\pi)^D}\frac{d^Dq}{(2\pi)^D}
 \frac{\mathcal{P}(\lam_i,p,q,k_i)}
{p^2q^2(p+q)^2(p-k_1)^2(q-k_2)^2(p+q+k_3)^2(p+q+k_3+k_4)^2},
\end{multline}
and
\begin{multline}
\mathcal{I}^\text{bow-tie}_4[\mathcal{P}(\lam_i,p,q,k_i)](s,t)
\\
= \int\frac{d^Dp}{(2\pi)^D}\frac{d^Dq}{(2\pi)^D}
\frac{\mathcal{P}(\lam_i,p,q,k_i)}
{p^2q^2(p-k_1)^2(p-k_1-k_2)^2(q-k_4)^2(q-k_3-k_4)^2},
\end{multline}
where the $k_i$ are the external momenta. The numerator factor
$\mathcal{P}(\lam_i,p,q,k_i)$ is a polynomial in the external and
loop momenta. The vectors $\lam_p$ and $\lam_q$ represent the
$(-2\eps)$-dimensional components of the loop momenta $p$ and $q$.
We use the notation $\lam_i^2=\lam_i\cdot\lam_i\geq0$ and
$\lam_{p+q}^2=(\lam_p+\lam_q)^2=\lam_p^2+\lam_q^2+2\lam_p\cdot\lam_q$.
The explicit values of these integrals, as a series in $\eps$ and
expressed in terms of polylogarithms, are given in appendix A of
ref.~\cite{Bern:2000dn}. We provide the bow-tie integrals in
eq.~\eqref{bowties} and the remaining ones in appendix \ref{sec:integrals}
of this manuscript. The symmetries obeyed by the color
factors~\eqref{adjcolorsym} carry over to the primitive
amplitudes~\eqref{gluprimamps}.
	
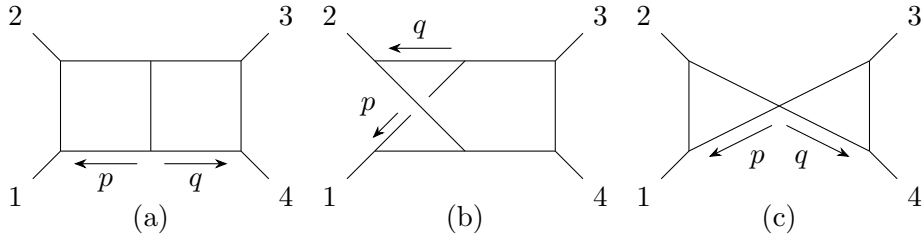
\begin{figure}[b]
		\centering
		\begin{tikzpicture}[scale=0.6]
			\begin{feynman}
				\vertex (g1) at (-3,-2) {\(1\)};
				\vertex (g2) at (-3,2) {\(2\)};
				\vertex (g3) at (3,2) {\(3\)};
				\vertex (g4) at (3,-2) {\(4\)};
				\vertex (v1) at (-2,-1);
				\vertex (v2) at (-2,1);
				\vertex (v3) at (2,1);
				\vertex (v4) at (2,-1);
				\vertex (v5) at (0,1);
				\vertex (v6) at (0,-1);
				\vertex (g6) at (0,-2.5) {(a)};
				\diagram*{
					(g1) -- (v1),
					(v2) -- (g2),
					(g3) -- (v3),
					(v4) -- (g4),
			(v1) -- [rmomentum'={[arrow distance=0.2]\(p\)}] (v6)
				-- [momentum'={[arrow distance=0.2]\(q\)}] (v4) 
					-- (v3) 
					-- (v5)
					-- (v2) 
					-- (v1),
					(v5) -- (v6)
					-- [opacity=0] (g6),
				};
			\end{feynman}
		\end{tikzpicture}
		\begin{tikzpicture}[scale=0.6]
			\begin{feynman}
				\vertex (g1) at (-3,-2) {\(1\)};
				\vertex (g2) at (-3,2) {\(2\)};
				\vertex (g3) at (3,2) {\(3\)};
				\vertex (g4) at (3,-2) {\(4\)};
				\vertex (v1) at (-2,-1);
				\vertex (v2) at (-2,1);
				\vertex (v3) at (2,1);
				\vertex (v4) at (2,-1);
				\vertex (v5) at (0,1);
				\vertex (v6) at (0,-1);
				\vertex (g6) at (0,-2.5) {(b)};
				\vertex (v7) at (-1.2,-0.2);
				\vertex (v8) at (-0.8,0.2);
				\diagram*{
					(g1) -- (v1),
					(v2) -- (g2),
					(g3) -- (v3),
					(v4) -- (g4),
					(v1) -- (v6)
					-- (v2) 
				-- [rmomentum={[arrow distance=0.2]\(q\)}] (v5)
					-- (v8),
			(v7) -- [momentum'={[arrow distance=0.3]\(p\)}] (v1),
					(v6) -- (v4)
					-- (v3) 
					-- (v5),
					(v6) -- [opacity=0] (g6),
				};
			\end{feynman}
		\end{tikzpicture}
		\begin{tikzpicture}[scale=0.6]
			\begin{feynman}
				\vertex (g1) at (-3,-2) {\(1\)};
				\vertex (g2) at (-3,2) {\(2\)};
				\vertex (g3) at (3,2) {\(3\)};
				\vertex (g4) at (3,-2) {\(4\)};
				\vertex (v1) at (-2,-1);
				\vertex (v2) at (-2,1);
				\vertex (v3) at (2,1);
				\vertex (v4) at (2,-1);
				\vertex (v6) at (0,0);
				\vertex (g6) at (0,-2.5) {(c)};
				\diagram*{
					(g1) -- (v1),
					(v2) -- (g2),
					(g3) -- (v3),
					(v4) -- (g4),
			(v1) -- [rmomentum'={[arrow distance=0.2]\(p\)}] (v6)
		  	-- [momentum'={[arrow distance=0.2]\(q\)}] (v4) 
					-- (v3) 
					-- (v6)
					-- (v2) 
					-- (v1),
					(v6) -- [opacity=0] (g6),
				};
\end{feynman}
\end{tikzpicture}
\caption{The three scalar integral topologies appearing in the two-loop
all-plus amplitude, with the loop-momentum routings displayed:
(a) the planar double box; (b) the non-planar double box; (c) the bow-tie.}
\label{fig:integrals}
\end{figure}

\subsection{Matter Contribution}
\label{sec:matter}

In order to compute the color factors for the fermionic matter contribution
in the representation $R_0$, one can simply replace the
fundamental loops appearing in the parent diagrams with a loop in $R_0$.
This replacement is allowed, because one can rewrite any color diagram
in terms of parent diagrams using only Jacobi identities, which
are independent of the fermion representation.
We denote the color factor given by a diagram $D_i$ with matter
representation $R_0$ by $R_{1234}^{D_i}$. The color decomposition
for the amplitude is then
\be
\label{Rmatter}
\mathcal{A}^{R_0}_F=g^6\sum_{D_i}\Big(
R_{1234}^{D_i}A_{F1234}^{D_i}+R_{3421}^{D_i}A_{F3421}^{D_i}  \Big)
  + \mathcal{C}(234),
\ee
where the seven parent diagrams $D_i$ are given in fig. \ref{fig:2loopF}.
The color factors $R^{D_i}$ are given by expanding the $R_0$ representation loops into a sum over the diagrams in fig.~\ref{fig:R1loop}, accounting for the relative minus sign that appears with a trace of an odd number of antifundamental generators. They are then evaluated in terms of traces in the fundamental with no contracted indices, using the rules given in eq.~\eqref{diagramrules}.

\begin{figure}
\centering
\begin{tikzpicture}[scale=0.6]
			\begin{feynman}
				\vertex (g1) at (-3.25,-2.25) {\(1\)};
				\vertex (g2) at (-3.25,2.25) {\(2\)};
				\vertex (g3) at (3.25,2.25) {\(3\)};
				\vertex (g4) at (3.25,-2.25) {\(4\)};
				\vertex (v1) at (-2,-1);
				\vertex (v2) at (-2,1);
				\vertex (v3) at (2,1);
				\vertex (v4) at (2,-1);
				\vertex (v5) at (0,1);
				\vertex (v6) at (0,-1);
				\vertex (g6) at (0,-2.5) {\(P_1\)};
				\diagram*{
					(g1) -- [gluon] (v1),
					(v2) -- [gluon] (g2),
					(g3) -- [gluon] (v3),
					(v4) -- [gluon] (g4),
					(v1) -- [anti fermion] (v6)
					-- [gluon] (v4) 
					-- [gluon] (v3) 
					-- [gluon] (v5)
					-- [anti fermion, edge label=\(R_0\),swap] (v2) 
					-- [anti fermion] (v1),
					(v5) -- [fermion] (v6)
					-- [opacity=0] (g6),
				};
			\end{feynman}
		\end{tikzpicture}
		\begin{tikzpicture}[scale=0.6]
			\begin{feynman}
				\vertex (g1) at (-3.25,-2.25) {\(1\)};
				\vertex (g2) at (-3.25,2.25) {\(2\)};
				\vertex (g3) at (3.25,2.25) {\(3\)};
				\vertex (g4) at (3.25,-2.25) {\(4\)};
				\vertex (v1) at (-2,-1);
				\vertex (v2) at (-2,1);
				\vertex (v3) at (2,1);
				\vertex (v4) at (2,-1);
				\vertex (v5) at (0,1);
				\vertex (v6) at (0,-1);
				\vertex (g6) at (0,-2.5) {\(P_2\)};
				\diagram*{
					(g1) -- [gluon] (v1),
					(v2) -- [gluon] (g2),
					(g3) -- [gluon] (v3),
					(v4) -- [gluon] (g4),
					(v1) -- [gluon] (v6)
					-- [anti fermion] (v4) 
					-- [anti fermion] (v3) 
					-- [anti fermion, edge label=\(R_0\), swap] (v5)
					-- [gluon] (v2) 
					-- [gluon] (v1),
					(v5) -- [anti fermion] (v6)
					-- [opacity=0] (g6),
				};
			\end{feynman}
		\end{tikzpicture}
		\begin{tikzpicture}[scale=0.6]
			\begin{feynman}
				\vertex (g1) at (-3.25,-2.25) {\(1\)};
				\vertex (g2) at (-3.25,2.25) {\(2\)};
				\vertex (g3) at (3.25,2.25) {\(3\)};
				\vertex (g4) at (3.25,-2.25) {\(4\)};
				\vertex (v1) at (-2,-1);
				\vertex (v2) at (-2,1);
				\vertex (v3) at (2,1);
				\vertex (v4) at (2,-1);
				\vertex (v5) at (0,1);
				\vertex (v6) at (0,-1);
				\vertex (g6) at (0,-2.5) {\(P_3\)};
				\diagram*{
					(g1) -- [gluon] (v1),
					(v2) -- [gluon] (g2),
					(g3) -- [gluon] (v3),
					(v4) -- [gluon] (g4),
					(v1) -- [anti fermion] (v6)
					-- [anti fermion] (v4) 
					-- [anti fermion] (v3) 
					-- [anti fermion, edge label=\(R_0\), swap] (v5)
					-- [anti fermion, edge label=\(R_0\), swap] (v2) 
					-- [anti fermion] (v1),
					(v5) -- [gluon] (v6)
					-- [opacity=0] (g6),
				};
			\end{feynman}
		\end{tikzpicture}
		\begin{tikzpicture}[scale=0.6]
			\begin{feynman}
				\vertex (g1) at (-3.75,-2.25) {\(1\)};
				\vertex (g2) at (-3.75,2.25) {\(2\)};
				\vertex (g3) at (3.75,2.25) {\(3\)};
				\vertex (g4) at (3.75,-2.25) {\(4\)};
				\vertex (v1) at (-2.5,-1);
				\vertex (v2) at (-2.5,1);
				\vertex (v3) at (2.5,1);
				\vertex (v4) at (2.5,-1);
				\vertex (v5) at (-0.75,0);
				\vertex (v6) at (0.75,0);
				\vertex (g6) at (0,-2.5) {\(P_4\)};
				\diagram*{
					(g1) -- [gluon] (v1),
					(v2) -- [gluon] (g2),
					(g3) -- [gluon] (v3),
					(v4) -- [gluon] (g4),
					(v1) -- [anti fermion] (v5)
					-- [anti fermion, edge label=\(R_0\), swap] (v2) 
					-- [anti fermion] (v1),
					(v4) -- [anti fermion] (v3)
					-- [anti fermion, edge label=\(R_0\), swap] (v6) 
					-- [anti fermion] (v4),
					(v6) -- [gluon] (v5),
					(v6) -- [opacity=0] (g6),
				};
			\end{feynman}
		\end{tikzpicture}
		\begin{tikzpicture}[scale=0.6]
			\begin{feynman}
				\vertex (g1) at (-3.25,-2.25) {\(1\)};
				\vertex (g2) at (-3.25,2.25) {\(2\)};
				\vertex (g3) at (3.25,2.25) {\(3\)};
				\vertex (g4) at (3.25,-2.25) {\(4\)};
				\vertex (v1) at (-2,-1);
				\vertex (v2) at (-2,1);
				\vertex (v3) at (2,1);
				\vertex (v4) at (2,-1);
				\vertex (v5) at (0,1);
				\vertex (v6) at (0,-1);
				\vertex (g6) at (0,-2.5) {\(NP_1\)};
				\vertex (v7) at (-1.2,-0.2);
				\vertex (v8) at (-0.8,0.2);
				\diagram*{
					(g1) -- [gluon] (v1),
					(v2) -- [gluon] (g2),
					(g3) -- [gluon] (v3),
					(v4) -- [gluon] (g4),
					(v8) -- [fermion] (v5)
					-- [gluon] (v2) 
					-- [gluon] (v6)
					-- [fermion] (v1),
					(v7) -- [anti fermion] (v1),
					(v6) -- [anti fermion] (v4)
					-- [anti fermion] (v3) 
					-- [anti fermion, edge label=\(R_0\), swap] (v5),
					(v6) -- [opacity=0] (g6),
				};
			\end{feynman}
		\end{tikzpicture}
		\begin{tikzpicture}[scale=0.6]
			\begin{feynman}
				\vertex (g1) at (-3.25,-2.25) {\(1\)};
				\vertex (g2) at (-3.25,2.25) {\(2\)};
				\vertex (g3) at (3.25,2.25) {\(3\)};
				\vertex (g4) at (3.25,-2.25) {\(4\)};
				\vertex (v1) at (-2,-1);
				\vertex (v2) at (-2,1);
				\vertex (v3) at (2,1);
				\vertex (v4) at (2,-1);
				\vertex (v5) at (0,1);
				\vertex (v6) at (0,-1);
				\vertex (g6) at (0,-2.5) {\(NP_2\)};
				\vertex (v7) at (-1.2,-0.2);
				\vertex (v8) at (-0.8,0.2);
				\diagram*{
					(g1) -- [gluon] (v1),
					(v2) -- [gluon] (g2),
					(g3) -- [gluon] (v3),
					(v4) -- [gluon] (g4),
					(v1) -- [gluon] (v6)
					-- [fermion] (v2) 
					-- [fermion] (v5)
					-- [gluon] (v8),
					(v7) -- [gluon] (v1),
					(v6) -- [anti fermion] (v4)
					-- [anti fermion] (v3) 
					-- [anti fermion, edge label=\(R_0\), swap] (v5),
					(v6) -- [opacity=0] (g6),
				};
			\end{feynman}
		\end{tikzpicture}
		\begin{tikzpicture}[scale=0.6]
			\begin{feynman}
				\vertex (g1) at (-3.25,-2.25) {\(1\)};
				\vertex (g2) at (-3.25,2.25) {\(2\)};
				\vertex (g3) at (3.25,2.25) {\(3\)};
				\vertex (g4) at (3.25,-2.25) {\(4\)};
				\vertex (v1) at (-2,-1);
				\vertex (v2) at (-2,1);
				\vertex (v3) at (2,1);
				\vertex (v4) at (2,-1);
				\vertex (v5) at (0,1);
				\vertex (v6) at (0,-1);
				\vertex (g6) at (0,-2.5) {\(NP_3\)};
				\vertex (v7) at (-1.2,-0.2);
				\vertex (v8) at (-0.8,0.2);
				\diagram*{
					(g1) -- [gluon] (v1),
					(v2) -- [gluon] (g2),
					(g3) -- [gluon] (v3),
					(v4) -- [gluon] (g4),
					(v1) -- [fermion] (v6)
					-- [fermion] (v2) 
					-- [fermion, edge label=\(R_0\)] (v5)
					-- [fermion] (v8),
					(v7) -- [fermion] (v1),
					(v6) -- [gluon] (v4)
					-- [gluon] (v3) 
					-- [gluon] (v5),
					(v6) -- [opacity=0] (g6),
				};
			\end{feynman}
		\end{tikzpicture}
\caption{Parent diagrams for the fermion loop $R_0$ contributions.}
\label{fig:2loopF}
\end{figure}
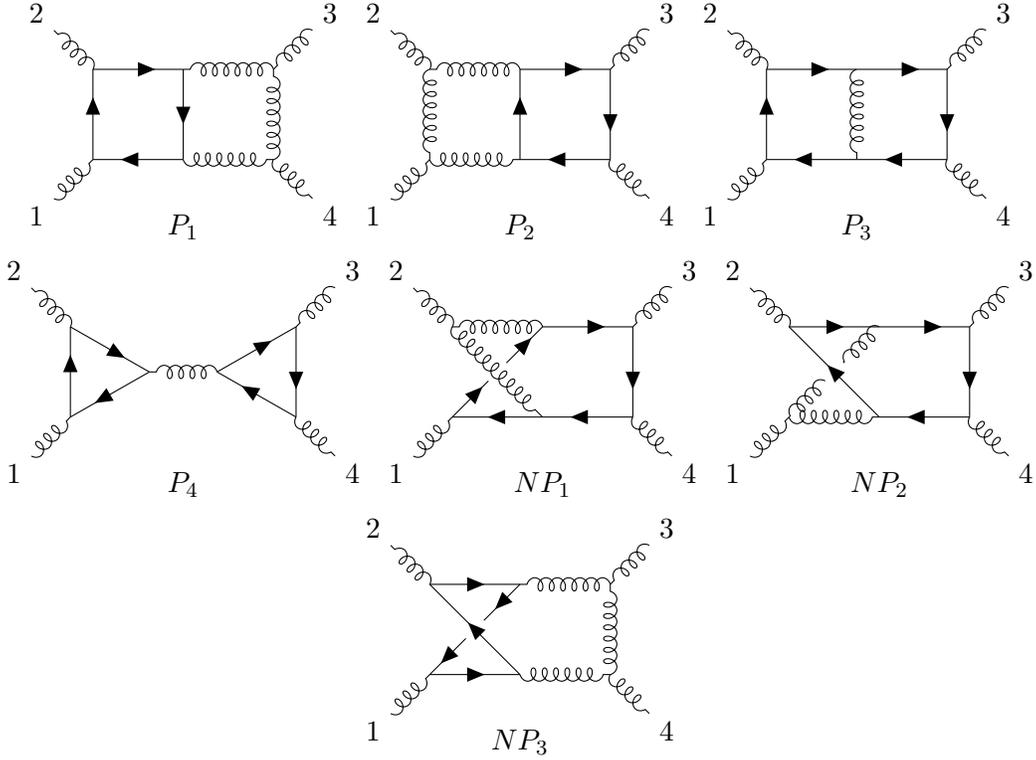

\begin{figure}
\centering
\be
\nonumber
			F^{P_4}_{1234}=
			\begin{tikzpicture}[baseline={(0,0)},scale=0.6]
				\begin{feynman}
					\vertex (g1) at (-3.75,-2.25) {\(1\)};
					\vertex (g2) at (-3.75,2.25) {\(2\)};
					\vertex (g3) at (3.75,2.25) {\(3\)};
					\vertex (g4) at (3.75,-2.25) {\(4\)};
					\vertex (v1) at (-2.5,-1);
					\vertex (v2) at (-2.5,1);
					\vertex (v3) at (2.5,1);
					\vertex (v4) at (2.5,-1);
					\vertex (v5) at (-0.75,0);
					\vertex (v6) at (0.75,0);
					\diagram*{
						(g1) -- [gluon] (v1),
						(v2) -- [gluon] (g2),
						(g3) -- [gluon] (v3),
						(v4) -- [gluon] (g4),
						(v1) -- [fermion] (v2)
						-- [fermion] (v5) 
						-- [fermion] (v1),
						(v4) -- [fermion] (v6)
						-- [fermion] (v3) 
						-- [fermion] (v4),
						(v6) -- [gluon] (v5),
						(v6) -- [opacity=0] (g6),
					};
				\end{feynman}
			\end{tikzpicture}
			+
			\begin{tikzpicture}[baseline={(0,0)},scale=0.6]
				\begin{feynman}
					\vertex (g1) at (-3.75,-2.25) {\(1\)};
					\vertex (g2) at (-3.75,2.25) {\(2\)};
					\vertex (g3) at (3.75,2.25) {\(3\)};
					\vertex (g4) at (3.75,-2.25) {\(4\)};
					\vertex (v1) at (-2.5,-1);
					\vertex (v2) at (-2.5,1);
					\vertex (v3) at (2.5,1);
					\vertex (v4) at (2.5,-1);
					\vertex (v5) at (-0.75,0);
					\vertex (v6) at (0.75,0);
					\diagram*{
						(g1) -- [gluon] (v1),
						(v2) -- [gluon] (g2),
						(g3) -- [gluon] (v3),
						(v4) -- [gluon] (g4),
						(v1) -- [fermion] (v5)
						-- [fermion] (v2) 
						-- [fermion] (v1),
						(v4) -- [fermion] (v3)
						-- [fermion] (v6) 
						-- [fermion] (v4),
						(v6) -- [gluon] (v5),
						(v6) -- [opacity=0] (g6),
					};
\end{feynman}
\end{tikzpicture}
\ee
\caption{The total contribution to $F^{P_4}_{1234}$. Notice that the
  terms $\tr_F(12c)\tr_{\bar{F}}(c34)$ and $\tr_{\bar{F}}(12c)\tr_F(c34)$
  with partially-reversed arrows do not contribute to its value.}
\label{fig:fp4}
\end{figure}
	
There is an exception to the evaluation procedure for $R_{1234}^{P_4}$,
since we follow the conventions of ref.~\cite{Bern:2002zk}.
In that reference, the color factors were evaluated by adding to
the diagram $P_4$ the contribution of the anti-fundamental representation
only, i.e.~they reverse the direction of the two arrows simultaneously.
In particular, the full diagrammatic color factor for $F_{1234}^{P_4}$ is
in fig.~\ref{fig:fp4}. Notice that the $F\times\bar{F}$ and
$\bar{F}\times F$ cross terms are not to be included;
their contributions are already included in the definition of
the kinematic factor $A_{F1234}^{P_4}$.
We must account for this convention by not including any terms of the form
$F\times\bar{F}$, $F\times\asyma$, $\asym\times\asyma$, and their
conjugates in $R_{1234}^{P_4}$. Thus, $R_{1234}^{P_4}$ is given by
\begin{align}
\begin{split}
R_{1234}^{P_4} =&~ 8^2\tr_F(12c)\tr_F(c34) + 8\tr_F(12c)\tr_\asym(c34) 
\\
&+ 8\tr_\asym(12c)\tr_F(c34) + \tr_\asym(12c)\tr_\asym(c34) 
\\
&+ \text{conjugate}.
\end{split}
\end{align}
	
With some help from trace identities provided in Appendix \ref{sec:colorid},
the results are
\begin{align}
\begin{split}
R_{1234}^{P_1}=&~(N^2+4N+2)\big[\tr(1234)+\tr(1432)\big] 
\\
&+ (-2N+2)\big[\tr(1243)+\tr(1342)\big] - 4\big[\tr(1324)+\tr(1423)\big]
\\
&+(6N+4)\tr(12)\tr(34) + 4\big[ \tr(13)\tr(24) + \tr(14)\tr(23) \big]\,,
\end{split}
\\
R_{1234}^{P_2} =&~ R_{1234}^{P_1} \,,
\\
\begin{split}
R_{1234}^{P_3} =&~ (N^2-6N+6+8N^{-1})\big[\tr(1234)+\tr(1432)\big]
\\
&+(-2N+6+8N^{-1})\big[\tr(1243)+\tr(1342)\big]
  + 8N^{-1}\big[\tr(1324)+\tr(1423)\big]
\\
&+(6N-8N^{-1})\tr(12)\tr(34) - 8N^{-1}\big[\tr(13)\tr(24)+\tr(14)\tr(23)\big]\,,
\end{split}
\\
\begin{split}
R_{1234}^{P_4} =&~ (N^2+10N+26)\big[\tr(1234)+\tr(1432)\big]
   +(-2N-10)\big[\tr(1243)+\tr(1342)\big]
\\
&+(-2N-16-32N^{-1})\tr(12)\tr(34),
\end{split}
\\
\begin{split}
R_{1234}^{NP_1}=&~2\big[\tr(1234)+\tr(1432)+\tr(1243)+\tr(1342)\big]
\\
&+ (2N-4)\big[\tr(1324)+\tr(1423)\big]
\\
&+ (4N+4)\tr(12)\tr(34) + (-2N+4)\big[\tr(13)\tr(24)+\tr(14)\tr(23)\big] \,,
\end{split}
\\
R_{1234}^{NP_2}=&~R_{1234}^{NP_1} \,,
\\
\begin{split}
R_{1234}^{NP_3}=&~(-2N+2)\big[\tr(1234)+\tr(1432)+\tr(1243)+\tr(1342)\big]
\\
&-4\big[\tr(1324)+\tr(1423)\big]
\\
&+(4N-8)\tr(12)\tr(34) + (-2N-8)\big[\tr(13)\tr(24)+\tr(14)\tr(23)\big] \,.
\end{split}
\end{align}
The color factors $R^{P_1}=R^{P_2}$ and $R^{NP_3}$ also have the same
symmetries as $C^P$ and $C^{NP}$, respectively, in eq.~\eqref{adjcolorsym}:
\begin{align}
\label{Rcolorsym}
\begin{split}
R^{P_1}_\text{1234} &= R^{P_1}_{3412} = R^{P_1}_{2143} = R^{P_1}_{4321} \,,
\\
R^{NP_3}_{1234} &= R^{NP_3}_{2134} = R^{NP_3}_{1243} = R^{NP_3}_{2143} \,.
\end{split}
\end{align}
Again, this is evident directly from the diagrams by applying rotations
and reflections to them, as well as using $R^{P_1}=R^{P_2}$.
	
The primitive amplitudes associated to each diagram $D_i$ are given in
ref.~\cite{Bern:2002zk}; however many of the integrals that compose them
vanish. Removing the integrals that vanish at ${\cal O}(\e^0)$,
the primitive amplitudes are
\begin{align}
\label{primamps}
\begin{split}
		A^{P_1}_{F1234} =&~ \rho
		\bigg\{
		s\mathcal{I}^P_4
		\big[
		-2\lam_p^2\lam_{p+q}^2
		\big]
		(s,t)
		\\
		&~~~~-2\frac{D_s-2}{s}\mathcal{I}_4^\text{bow-tie}
		\Big[
		\lam_p^2\lam_q^2
		\big(
		(p+q)^2+s
		\big)
		\Big]
		(s,t)
		\bigg\},
	\\
		A^{P_2}_{F1234} =&~ \rho
		\bigg\{
		s\mathcal{I}^P_4
		\big[
		-2\lam_q^2\lam_{p+q}^2
		\big]
		(s,t)
		\\
		&~~~~-2\frac{D_s-2}{s}\mathcal{I}_4^\text{bow-tie}
		\Big[
		\lam_p^2\lam_q^2
		\big(
		(p+q)^2+s
		\big)
		\Big]
		(s,t)
		\bigg\},
	\\
		A^{P_3}_{F1234} =&~ \rho
		(D_s-2)\mathcal{I}_4^\text{bow-tie}
		\big[
		\lam_p^2\lam_q^2
		\big]
		(s,t),
	\\
		A^{P_4}_{F1234} =&~ \rho
		\frac{4}{s}\mathcal{I}_4^\text{bow-tie}
		\Big[
		\lam_p^2\lam_q^2
		\big(
		(p+q)^2+\tfrac{1}{2}s
		\big)
		\Big]
		(s,t), 
	\\
		A^{NP_1}_{F1234} =&~\rho s \mathcal{I}^{NP}_4
		\big[
		-2\lam_p^2\lam_{p+q}^2
		\big](s,t),
	\\
		A^{NP_2}_{F1234} =&~\rho s \mathcal{I}^{NP}_4
		\big[
		-2\lam_q^2\lam_{p+q}^2
		\big](s,t),
	\\
		A^{NP_3}_{F1234} =&~\rho s \mathcal{I}^{NP}_4
		\big[
		-2\lam_p^2\lam_q^2
		\big](s,t).
\end{split}
\end{align}
The bow-tie integrals are quite simple, as they are products of one-loop
triangle integrals, and are given by~\cite{Bern:2000dn}
\begin{align}
\label{bowties}
\begin{split}
		\mathcal{I}_4^\text{bow-tie}[\lam_p^2\lam_q^2](s,t) 
		&= -\frac{1}{4}\frac{1}{(4\pi)^4},
		\\
		\mathcal{I}_4^\text{bow-tie}[\lam_p^2\lam_q^2(p+q)^2](s,t)
		&= -\frac{1}{36}\frac{1}{(4\pi)^4}(t-4s).
\end{split}
\end{align}
We provide the results for the remaining integrals from
ref.~\cite{Bern:2000dn} in appendix \ref{sec:integrals}.
The primitive amplitudes $A^{P_1}$, $A^{P_2}$, and $A^{NP_3}$ obey the
same relations as their corresponding color factors in eq.~\eqref{Rcolorsym}.
For later use, notice that these amplitudes are the only ones out of the
matter contribution that contain $1/\eps$ poles.

\subsection{The full amplitude}
\label{sec:full}

The two-loop four-gluon amplitude with fermionic matter in the
representation $R_0$ is the sum of eqs.~\eqref{Rmatter} and \eqref{pureglu}.
Using the values of the color factors given above, we have
\begin{align}
\begin{split}
  \mathcal{A}^\text{2-loop}_4 =&~ g^6\sum_{\sigma\in S_4/\mathbb{Z}_4}
  A_{4;1}^\text{2-loop}(\sigma(1,2,3,4))\tr(\sigma(1,2,3,4))
\\
&+ g^6\sum_{\sigma\in S_4/S_{4;3}} A_{4;3}^\text{2-loop}(\sigma(1,2,3,4))
                              \tr(\sigma(1,2))\tr(\sigma(3,4)),
\end{split}
\end{align}
where the $A_{4;c}^\text{2-loop}$, $c=1,3$
are the two-loop color-ordered subamplitudes.
The $A_{4;c}$ contain various powers of $N$, so we separate them
into these different powers as
\be
A_{4;c}^\text{2-loop} = A_{4;c;2}N^2 + A_{4;c;1}N + A_{4;c;0} + A_{4;c;-1}N^{-1}.
\ee
Of course, the $A_{4;c;i}$ are linear combinations of the primitive
amplitudes $A_{G1234}^{D_i}$ and $A_{F1234}^{D_i}$, which are given by
\begin{align}
  A_{4;1;2}(1,2,3,4) =&~ A_{G1234}^P + A_{G2341}^P
        + \sum_{i=1}^{4}\big(A_{F1234}^{P_i}+A_{F2341}^{P_i}\big),
\\
\begin{split}
  A_{4;1;1}(1,2,3,4) =&~ 2A_{F1234}^Z
   + 2A_{F2341}^Z - 2A_{F1234}^{NP_3} - 2A_{F2341}^{NP_3}
\\
&-2A_{F3421}^{NP_3}-2A_{F1423}^{NP_3}
-2\sum_{i=1}^{4}\big( A_{F3421}^{P_i} + A_{F1423}^{P_i} \big)
\\
&+2A_{F1342}^{NP_1} + 2A_{F1342}^{NP_2} + 2A_{F4231}^{NP_1} + 2A_{F4231}^{NP_2},
\end{split}
\\
\begin{split}
\label{N0}
A_{4;1;0}(1,2,3,4) =&~ 2A_{1234}^\text{all}
+2A_{3421}^\text{all}-4A_{1342}^\text{all}
-4A_{4231}^\text{all}+2A_{1423}^\text{all}+2A_{2341}^\text{all}
\\
&+4\big(A_{F1234}^{P_3} + \text{perms}\big)
\\
&+24A_{F1234}^{P_4} - 12A_{F3421}^{P_4} + 4A_{F1342}^{P_4}
\\
&+ 4A_{F4231}^{P_4} -12A_{F1423}^{P_4} + 24A_{F2341}^{P_4},
\end{split}
\\
A_{4;1;-1}(1,2,3,4) =&~ 8\big(A_{F1234}^{P_3} + \text{perms}\big),
\\
A_{4;3;2}(1,2,3,4) =&~ 0,
\\
\begin{split}
A_{4;3;1}(1,2,3,4) =&~ 6A_{1234}^\text{all} + 6A_{3421}^\text{all}
        -6A_{F1234}^{P_4} - 6A_{F3421}^{P_4}
\\
&-2\big(
A_{G1234}^{NP}+\sum_{i=1}^{3}A_{F1234}^{NP_i}
		+ \text{perms}
		\big),
\end{split}
\\
\begin{split}
A_{4;3;0}(1,2,3,4) =&~ 4\big(
A_{F1234}^{P_1} + A_{F1234}^{P_2} + A_{F1234}^{NP_1} + A_{F1234}^{NP_2} 
 - 2A_{F1234}^{NP_3} + \text{perms}
		\big)
\\
		&-16A_{F1234}^{P_4} - 16A_{F3421}^{P_4},
\end{split}
\\
\begin{split}
A_{4;3;-1}(1,2,3,4) =&~ -8\big(A_{F1234}^{P_3} + \text{perms}\big)
  -32A_{F1234}^{P_4} - 32A_{F3421}^{P_4},
\end{split}
\end{align}
where $A_{1234}^\text{all}=A_{G1234}^P+A_{G1234}^{NP} + \sum_{D_i}A_{F1234}^{D_i}$,
the sum of all 9 primitive amplitudes, and 
\be
A_{F1234}^Z \equiv 2A_{F1234}^{P_1} + 2A_{F1234}^{P_2} 
 - 3A_{F1234}^{P_3} + 5A_{F1234}^{P_4} \,.
\ee
The term ``$+\text{perms}$'' means to add all non-trivial permutations
\be
 (3,4,2,1),\, (1,3,4,2), \, (4,2,3,1), \, (1,4,2,3), \, (2,3,4,1),
\ee
of the preceding terms inside the parentheses.
	 
\subsection{Dimensional regularization scheme}
\label{sec:dimreg}

The primitive amplitudes are evaluated in
refs.~\cite{Bern:2000dn,Bern:2002zk} using dimensional regularization
with the loop momentum being in $D=4-2\epsilon>4$ dimensions.
The dimension of the ``unobserved" internal gluonic states $D_s$ is left
explicit in their results, with $D_s\geq D$ in intermediate steps of their
calculation. The unobserved states include virtual states in loops
and virtual intermediate states in trees. Setting $D_s=D$ corresponds to
the standard 't Hooft-Veltman (HV) scheme. In the four-dimensional
helicity (FDH) scheme, one would set $D_s=4$.
	
The choice of $D_s$ affects the compliance of the amplitudes with
supersymmetry Ward identities
(SWI)~\cite{Grisaru:1976vm,Grisaru:1977px,Parke:1985pn}.
In particular, preserving the
number of bosonic states relative to the fermionic states is necessary
for preserving the SWI, and the choice $D_s=4$ achieves
this~\cite{Bern:2002zk}. The SWI manifest themselves in terms of the
primitive amplitudes as
\begin{align}
\label{SWI}
\begin{split}
A_{G1234}^P+\sum_{i=1}^{4}A_{F1234}^{P_i}&=0,
\\
A_{G1234}^{NP}+\sum_{i=1}^{3}A_{F1234}^{NP_i}&=0
\end{split}
\end{align}
in the $\epsilon\to0$ limit. These identities do not hold in the HV scheme.
Applying the constraints in eq.~\eqref{SWI} simplifies the $A_{4;c;i}$
considerably, and we get agreement with eq.~\eqref{KCformula} at order $N^2$.
Moreover, the choice $D_s=4$ forces the one-loop partial amplitudes
$A_n^{[1]}$ and $A_n^{[1/2]}$ to be equal with opposite signs to all orders
in $\eps$. For these reasons, we take $D_s=4$ when evaluating the linear
combinations of primitive amplitudes.
	
When eq.~\eqref{SWI} is applied, the expressions for the $A_{4;c;i}$ in
terms of the primitive amplitudes simplify to
\begin{align}
A_{4;1;2}(1,2,3,4) &= 0,
\\
\label{A411}
\begin{split}
A_{4;1;1}(1,2,3,4) &= 2A_{F1234}^Z
  + 2A_{F2341}^Z - 2A_{F1234}^{NP_3} - 2A_{F2341}^{NP_3}
\\
&-2A_{F3421}^{NP_3}-2A_{F1423}^{NP_3}
  -2\sum_{i=1}^{4}\big( A_{F3421}^{P_i} + A_{F1423}^{P_i} \big)
\\
&+2A_{F1342}^{NP_1} + 2A_{F1342}^{NP_2} + 2A_{F4231}^{NP_1} + 2A_{F4231}^{NP_2},
\end{split}
\\
A_{4;1;0}(1,2,3,4) &= 4\big(A_{F1234}^{P_3} +\text{perms}\big) 
			+ 36A_{F1234}^{P_4} + 36A_{F2341}^{P_4},
\\
A_{4;1;-1}(1,2,3,4) &= 8\big(A_{F1234}^{P_3} + \text{perms}\big),
\\
A_{4;3;2}(1,2,3,4) &= 0,
\\
A_{4;3;1}(1,2,3,4) &= 0,
\\
\label{A430}
A_{4;3;0}(1,2,3,4) &=
4 \big( A_{F1234}^{P_1} + A_{F1234}^{P_2} + A_{F1234}^{NP_1}
  + A_{F1234}^{NP_2} - 2A_{F1234}^{NP_3} + \text{perms} \big),
\\
A_{4;3;-1}(1,2,3,4) &= -8\big( A_{F1234}^{P_3} + \text{perms} \big),
\end{align}
where we also made use of the fact that $A_{F1234}^{P_4}=-A_{F3421}^{P_4}$,
which follows from the reflection identity of the associated color
factor, $F_{1234}^{P_4}=-F_{3421}^{P_4}$. 

\subsection{Evaluating the $A_{4;c;i}$}
\label{sec:EvalA4ci}

The vanishing of $A_{4;1;2}$, $A_{4;3;2}$, and $A_{4;3;1}$ agrees with
eq.~\eqref{KCformula}. Using the expressions given
in eqs.~\eqref{primamps} and \eqref{bowties}, we see that
$A_{4;1;0}$, $A_{4;1;-1}$, and $A_{4;3;-1}$ are finite and rational.
They evaluate to
\begin{align}
A_{4;1;0}(1,2,3,4) &= -\frac{4\rho}{(4\pi)^4}\frac{s^2+4st+t^2}{st} \,,
\\
A_{4;1;-1}(1,2,3,4) &= -\frac{24\rho}{(4\pi)^4} \,,
\\
A_{4;3;-1}(1,2,3,4) &= \frac{24\rho}{(4\pi)^4} \,,
\end{align}
agreeing with eq.~\eqref{KCformula} at the corresponding powers of $N$.
	
The two remaining linear combinations of the primitive amplitudes,
$A_{4;1;1}$ and $A_{4;3;0}$, do not simplify further using the SWI,
and they explicitly contain non-finite and non-rational primitive
amplitudes (see appendix \ref{sec:integrals}).
Schematically, they are of the form
\be
\label{A411func}
A_{4;1;1}(1,2,3,4)=\frac{12\rho}{(4\pi)^4} 
+ \frac{1}{\eps}\text{transcendental} + \text{transcendental}
+ {\cal O}(\eps),
\ee
and
\be
\label{A430func}
A_{4;3;0}(1,2,3,4) = \frac{24\rho}{(4\pi)^4} + \text{transcendental}
+ {\cal O}(\eps),
\ee
where transcendental refers to terms that
(after multiplying by $(4\pi)^4$) contain products of
$\ln$, $\Li_2$, and $\Li_3$, which have as their arguments
$\pm t/s$, $1+t/s$, and $t/(s+t)$
and which have rational coefficients in $t/s$.
	
These expressions clearly do not agree with eq.~\eqref{KCformula},
because they have transcendental terms and/or $1/\eps$ poles,
along with the rational terms shown explicitly, which
do appear in eq.~\eqref{KCformula}.
This might at first seem to invalidate eq.~\eqref{KCformula}.
However, a comparison of the whole amplitude with the expected universal
IR behavior of two-loop amplitudes given by \cite{Catani:1998bh} sheds
light on the matter.  We carry out this comparison next.

	
\section{IR subtraction}
\label{sec:IRsubtraction}

In this section, we compare the IR behavior of eqs.~\eqref{A411func}
and \eqref{A430func} to that predicted on general principles.
We follow closely the analysis in section~5 of ref.~\cite{Bern:2000dn}.
The principal issue is that the IR behavior of a two-loop amplitude
in dimensional regularization involves $1/\e^2$ poles multiplying
the one-loop amplitude, so that higher order terms in $\e$ are required.
And while the one-loop amplitude in our case vanishes at ${\cal O}(\e^0)$,
it does \emph{not} vanish at higher orders in $\e$, because the
box integrals that enter it do not have the same symmetry properties
beyond leading order in $\e$.

Catani provides a universal factorization formula for dimensionally regulated,
UV-renorm\-alized two-loop amplitudes~\cite{Catani:1998bh}.
In the color-space operator formalism, the renormalized two-loop $n$-point
amplitude is given by
\begin{align}
\label{Catani}
\begin{split}
|\mathcal{M}_n^{(2)}(\mu^2;\{p\})\rangle_\text{R.S.}
=~ &\mathbf{I}^{(1)}(\epsilon,\mu^2;\{p\})
 |\mathcal{M}_n^{(1)}(\mu^2;\{p\})\rangle_\text{R.S.}
\\
+&~\mathbf{I}^{(2)}_\text{R.S.}(\epsilon,\mu^2;\{p\})
|\mathcal{M}_n^{(0)}(\mu^2;\{p\})\rangle_\text{R.S.}
  + |\mathcal{M}_n^{(2),\text{fin}}(\mu^2;\{p\})\rangle_\text{R.S.},
\end{split}
\end{align}
where $|\mathcal{M}_n^{(L)}(\mu^2;\{p\})\rangle_\text{R.S.}$ is the vector in
color space that represents the renormalized $L$-loop amplitude.
The subscript R.S. signifies a dependence on the renormalization scheme,
and $\mu$ is the renormalization mass scale.
For notational simplicity, we set $\mu=1$. The amplitudes are recovered by
\be
\mathcal{A}_n(1^{a_1},\dotsc,n^{a_n})
= \langle a_1,\dotsc,a_n|\mathcal{M}_n(p_1,\dotsc,p_n)\rangle,
\ee
where the $a_i$ is the color index of the $i$-th external parton.

The operators $\mathbf{I}^{(1)}$ and $\mathbf{I}^{(2)}$ encode the IR
divergences of $\mathcal{A}_n$. For all-plus helicity external gluons,
the tree-level amplitude $\mathcal{M}_n^{(0)}(\mu^2;\{p\})$ vanishes,
meaning that only $\mathbf{I}^{(1)}$ contributes to the divergences.
This operator is given by
\be
\label{Iop}
\mathbf{I}^{(1)}(\epsilon;\{p\}) = \frac{c_\Gamma}{2}
\sum_{i=1}^{n}\sum_{j\neq i}^{n}\mathbf{T}_i\cdot\mathbf{T}_j
\Bigg[
\frac{1}{\epsilon^2}
\bigg(\frac{e^{-i\lambda_{ij}\pi}}{s_{ij}}
\bigg)^\epsilon + 2\frac{\gamma_i}{\mathbf{T}_i^2}\frac{1}{\epsilon}
\Bigg] \,,
\ee
where $\lambda_{ij}=+1$ if $i$ and $j$ are both incoming or outgoing partons
and $\lambda_{ij}=0$ otherwise. The factor $c_\Gamma$ is
\be
c_\Gamma = \frac{1}{(4\pi)^{2-\eps}}
\frac{\Gamma(1+\eps)\Gamma^2(1-\eps)}{\Gamma(1-2\eps)} \,.
\ee
The color charge $\mathbf{T}_i=\{T_i^a\}$ is a vector with respect to the
generator label $a$ and an $SU(N)$ matrix  with respect to the color indices
of the outgoing parton $i$. For the adjoint representation $T^a_{bc}=if^{bac}$,
so $\mathbf{T}_i^2=C_A=2T_FN$.

For external gluons, $\gamma_i=b_0$,
where $b_0$ is the one-loop $\beta$-function coefficient.
For QCD with $N_F$ quark flavors,
\be
b_0^\text{QCD} = \frac{11C_A-4T_FN_F}{6} \,,
\ee
and for fermions in the representation $R_0$ (see eq.~\eqref{Tasym}),
\be
b_0^{R_0} = \frac{11C_A-4T_FN_F-4T_\asym N_\asym}{6}
\bigg|_{T_F=1,N_F=8,T_\asym=N-2,N_\asym=1}
= 3N-4.
\ee
Note that eq.~\eqref{Iop} differs slightly from Catani's original formula.
We have defined our structure constants such that they are greater by a
factor of $\sqrt{2}$, and we have included a factor of $2c_\Gamma$ instead of
$e^{\epsilon\gamma}$ due to a different normalization convention
for the coupling expansion parameter
($g^2$ vs. Catani's $\alpha_s/(2\pi) = g^2/(8\pi^2)$).

For external gluons of positive helicity only, we can rewrite the
predicted divergent part of the renormalized two-loop amplitude
in our notation as
\be
\label{Aijsum}
\mathcal{A}_n^\text{2-loop,ren.}(1^{a_1},\dotsc,n^{a_n})
\Big|_\text{pred. div.} = 
g^2 \sum_{1\leq i<j\leq n}\mathcal{A}_n^{(i,j)}(1,\dotsc,n),
\ee
where 
\begin{align}
\label{Aij}
\begin{split}
\mathcal{A}_n^{(i,j)}(1,\dotsc,n)\ =\ c_\Gamma & (if^{a_icb_i})(if^{a_jcb_j})
\bigg[
\frac{1}{\epsilon^2}(-s_{ij})^{-\epsilon}+2\frac{b_0^{R_0}}{C_A}\frac{1}{\epsilon}
\bigg]
\\
&\times
\mathcal{A}_{n}^\text{1-loop}(1^{a_1},\dotsc,i^{b_i},\dotsc,j^{b_j},\dotsc,n^{a_n})
\end{split}
\end{align}
acts on the colors of legs $i$ and $j$.
Specializing to four points, we only need to evaluate the case $(i,j)=(1,2)$,
\begin{align}
\label{A12}
\begin{split}
\mathcal{A}_4^{(1,2)}(1^{a_1},2^{a_2},3^{a_3},4^{a_4})
= & c_\Gamma (if^{a_1cb_1})(if^{a_2cb_2})
\bigg[
\frac{1}{\epsilon^2}(-s)^{-\epsilon}+2\frac{b_0^{R_0}}{C_A}\frac{1}{\epsilon}
\bigg]
\\
&\times
\mathcal{A}_4^\text{1-loop}(1^{b_1},2^{b_2},3^{a_3},4^{a_4}),
\end{split}
\end{align}
as the other five cases are obtained by relabeling $i$ and $j$.
	
It should be noted that there are two conventions for the placement of
$(e^{-i\lambda_{ij}\pi}/s_{ij})^{\eps}$ in eq.~\eqref{Iop}. The other convention is
to have it multiplying both powers of $\eps$ rather than just the
$\eps^{-2}$ as in eq.~\eqref{Iop}. With our choice of the matter representation,
the two choices are equivalent up to and including
${\cal O}(\eps^0)$,
since the one-loop amplitude vanishes identically at $\eps^0$, i.e. 
\be
\bigg[
\frac{1}{\eps^2}(-s_{ij})^{-\eps} + 2\frac{b_0}{C_A}\frac{1}{\eps}
\bigg]
\mathcal{A}_n^\text{1-loop}
=
\bigg[
\frac{1}{\eps^2} + 2\frac{b_0}{C_A}\frac{1}{\eps}
\bigg]
(-s_{ij})^{-\eps}
\mathcal{A}_n^\text{1-loop}
+ {\cal O}(\eps).
\ee
	
The one-loop amplitude with matter in the representation $R_0$ decomposes as
\begin{align}
\label{A1loop}
\begin{split}
\mathcal{A}_4^\text{1-loop}(1,2,3,4) = g^4
&\Big[
C_{G1234}^\text{1-loop}A_4^{[1]}(1,2,3,4)
+ C_{G1243}^\text{1-loop}A_4^{[1]}(1,2,4,3)
+ C_{G1423}^\text{1-loop}A_4^{[1]}(1,4,2,3)
\\
&\hskip-0.7cm + C_{R_01234}^\text{1-loop}A_4^{[1/2]}(1,2,3,4)
+ C_{R_01243}^\text{1-loop}A_4^{[1/2]}(1,2,4,3)
+ C_{R_01423}^\text{1-loop}A_4^{[1/2]}(1,4,2,3)
\Big].
\end{split}
\end{align}
Here, the $C_{X1234}^\text{1-loop}$ with $X\in\{G,R_0\}$ are given by ring graphs
with the loop being in the representation $X$. They are depicted in the
left-hand side of fig.~\ref{fig:g=ffbar} for $X=G$ and
in fig.~\ref{fig:R1loop} for $X=R_0$.

The kinematic factors $A_4^{[j]}$ are the familiar one-loop color-ordered
all-plus amplitudes for a particle of spin $j$ propagating in the loop.
However, unlike in eq.~\eqref{1looppartials}, here we will need the
result to higher orders in $\e$:
\begin{align}
A_4^{[1]}(1,2,3,4) &=
-(D_s-2)i\rho\mathcal{I}_4^\text{1-loop}[\lambda_\ell^4](s,t),
\\
A_4^{[1/2]}(1,2,3,4) &= 2i\rho\mathcal{I}_4^\text{1-loop}[\lambda_\ell^4](s,t),
\end{align}
with
\be
\mathcal{I}_4^\text{1-loop}[\lambda_\ell^4](s,t)
= \int\frac{d^D\ell}{(2\pi)^D}
\frac{\lambda_\ell^4}{\ell^2(\ell-k_1)^2(\ell-k_1-k_2)^2(\ell+k_4)^2} \,,
\label{finitebox}
\ee
and $\lambda_\ell$ represents the $(-2\e)$-dimensional components
of the loop momentum $\ell$.
The box integral $\mathcal{I}_4^\text{1-loop}[\lambda_\ell^4]$ is finite
as $\epsilon\to0$ so that 
\be
A_4^{[1/2]}(1,2,3,4) = - A_4^{[1]}(1,2,3,4),
\ee
in this limit, or when $D_s=4$.
We will keep $A_4^{[1]}$ and $A_4^{[1/2]}$ distinct for now.
	
After inserting eq.~\eqref{A1loop} into eq.~\eqref{A12},
the two structure constants from the operator $\mathbf{I}^{(1)}$
will be contracted with the different one-loop color coefficients,
and these contractions give rise to two-loop color diagrams,
\begin{align}
\begin{split}
(if^{b_1a_1c})(if^{ca_2b_2})C_{Gb_1b_234}^\text{1-loop} &= C^P_{1234},
\\
(if^{b_1a_1c})(if^{ca_2b_2})C_{Gb_1b_243}^\text{1-loop} &= C^P_{1243},
\\
(if^{b_1a_1c})(if^{ca_2b_2})C_{Gb_14b_23}^\text{1-loop} &= C^{NP}_{3412},
\\
(if^{b_1a_1c})(if^{ca_2b_2})C_{R_0b_1b_234}^\text{1-loop} &= R^{P_1}_{1234},
\\
(if^{b_1a_1c})(if^{ca_2b_2})C_{R_0b_1b_243}^\text{1-loop} &= R^{P_1}_{1243},
\\
(if^{b_1a_1c})(if^{ca_2b_2})C_{R_0b_14b_23}^\text{1-loop} &= R^{NP_3}_{3412}.
\end{split}
\end{align}
These relations allow us to write $\mathcal{A}_4^{(1,2)}$ as
\begin{align}
\label{A12R}
\begin{split}
\mathcal{A}_4^{(1,2)}(1,2,3,4) = - g^6
&\bigg[
\frac{1}{\epsilon^2}(-s)^{-\epsilon}+2\frac{b_0^{R_0}}{C_A}\frac{1}{\epsilon}
\bigg]
\\
& \hskip-0.5cm \times
\Big[
C^P_{1234} A_4^{[1]}(1,2,3,4)
+ C^P_{1243} A_4^{[1]}(1,2,4,3)
+ C^{NP}_{3412} A_4^{[1]}(1,3,2,4)
\\
& \hskip-0.5cm
+ R^{P_1}_{1234} A_4^{[1/2]}(1,2,3,4)
+ R^{P_1}_{1243} A_4^{[1/2]}(1,2,4,3)
+ R^{NP_3}_{3412} A_4^{[1/2]}(1,3,2,4)
\Big].
\end{split}
\end{align}
Now we insert eq.~\eqref{A12R} into eq.~\eqref{Aijsum} and perform the sum
over $i$ and $j$ by first adding the term with $(i,j)=(3,4)$.
We arrive at
\begin{align}
\label{divpred}
\begin{split}
\mathcal{A}_n^\text{2-loop,ren.}(1,2,3,4)
\Big|_\text{pred. div.} =&~ -g^6 c_\Gamma
\bigg[
\frac{1}{\epsilon^2}(-s)^{-\epsilon}+2\frac{b_0^{R_0}}{C_A}\frac{1}{\epsilon}
\bigg]
\\
&\times\Big[
2 C^P_{1234} A_4^{[1]}(1,2,3,4) + 2 C^P_{1243} A_4^{[1]}(1,2,4,3)
\\
&\hspace{0.5cm}
+ \big( C^{NP}_{3412} + C^{NP}_{1234} \big) A_4^{[1]}(1,3,2,4)
\\
&\hspace{0.5cm}
+ 2 R^{P_1}_{1234} A_4^{[1/2]}(1,2,3,4) + 2 R^{P_1}_{1243} A_4^{[1/2]}(1,2,4,3)
\\
&\hspace{0.5cm}
+ \big( R^{NP_3}_{3412} + R^{NP_3}_{1234} \big) A_4^{[1/2]}(1,3,2,4)
\Big]
\\
&+ \mathcal{C}(234).
\end{split}
\end{align}
	
Let us compare the predicted two-loop divergences for matter in the
representation $R_0$ eq.~\eqref{divpred} to those appearing in the
actual two-loop amplitude. There are two divergent integrals contributing
to the this amplitude, namely
$\mathcal{I}_4^P[\lambda_q^2\lambda_{p+q}^2]
=\mathcal{I}_4^P[\lambda_p^2\lambda_{p+q}^2]$
and $\mathcal{I}_4^{NP}[\lambda_p^2\lambda_q^2]$ \cite{Bern:2000dn}.
The divergent parts of these integrals are proportional to the
one-loop box integral,
\begin{align}
\label{intdivpart}
\begin{split}
\mathcal{I}_4^P[\lambda_q^2\lambda_{p+q}^2](s,t)\Big|_\text{div.}
&= -ic_\Gamma \, \frac{1}{\epsilon^2}(-s)^{-1-\epsilon}
 \, \mathcal{I}_4^\text{1-loop}[\lambda_\ell^4](s,t),
\\
\mathcal{I}_4^{NP}[\lambda_p^2\lambda_q^2](s,t)\Big|_\text{div.}
&= -ic_\Gamma \, \frac{1}{\epsilon^2}(-s)^{-1-\epsilon}
 \, \mathcal{I}_4^\text{1-loop}[\lambda_\ell^4](u,t),
\end{split}
\end{align}
as expected if eq.~\eqref{divpred} is to be recovered.
A heuristic reason for this factorization is given in ref.~\cite{Bern:2000dn},
but we briefly summarize it here. When loop momenta are simultaneously
soft and collinear with two adjacent external legs,
three consecutive propagators can go on shell. When they go on shell,
the remaining propagators become exactly that of the finite box integral
with external momenta $k_1,k_2,k_3,k_4$ in the planar case and
$k_1,k_4,k_2,k_3$ in the nonplanar case. The $(-2\e)$-dimensional numerator
in both cases becomes the numerator $\lambda_\ell^4$ in
eq.~\eqref{finitebox}.
The spacetime picture is then a small finite box
times an enlarged divergent triangle.

The divergences of the primitive amplitudes in terms of the one-loop
amplitudes are given by
\begin{align}
\begin{split}
A^P_{G1234}\Big|_\text{div.}&=-2c_\Gamma\frac{1}{\epsilon^2}
(-s)^{-\epsilon} A_4^{[1]}(1,2,3,4),
\\
A^{NP}_{G1234}\Big|_\text{div.}&=-c_\Gamma\frac{1}{\epsilon^2}
(-s)^{-\epsilon} A_4^{[1]}(1,3,2,4),
\\
A^{P_1}_{F1234}\Big|_\text{div.}
= A^{P_2}_{F1234}\Big|_\text{div.}&=-c_\Gamma\frac{1}{\epsilon^2}
(-s)^{-\epsilon} A_4^{[1/2]}(1,2,3,4),
\\
A^{NP_3}_{F1234}\Big|_\text{div.}&=-c_\Gamma\frac{1}{\epsilon^2}
(-s)^{-\epsilon} A_4^{[1/2]}(1,3,2,4),
\\
A^{D_i}_{F1234}\Big|_\text{div.}&=0,
\end{split}
\end{align}
where $D_i\in\{P_3,P_4,NP_1,NP_2\}$ for the last equality.
Plugging these formulae
into the sum of \eqref{Rmatter} and \eqref{pureglu} yields
\begin{align}
\begin{split}
\mathcal{A}_4^\text{2-loop}(1,2,3,4)\Big|_\text{div.}
&= g^6\Big[
C^P_{1234}A^P_{G1234} + C^P_{3421}A^P_{G3421} 
+ C^{NP}_{1234}A^{NP}_{G1234} + C^{NP}_{3421}A^{NP}_{G3421} 
\\
&\hspace{1cm}+ R^{P_1}_{1234}A^{P_1}_{F1234} + R^{P_1}_{3421}A^{P_1}_{F3421}
+ R^{P_2}_{1234}A^{P_2}_{F1234} + R^{P_2}_{3421}A^{P_2}_{F3421}
\\
&\hspace{1cm}+ R^{NP_3}_{1234}A^{NP_3}_{F1234} + R^{NP_3}_{3421}A^{NP_3}_{F3421}
\Big]\Big|_\text{div.}
+ \mathcal{C}(234)
\\
&=- g^6 c_\Gamma \frac{1}{\epsilon^2} (-s)^{-\epsilon}
\Big[
2 C^P_{1234} A_4^{[1]}(1,2,3,4) + 2 C^P_{1243} A_4^{[1]}(1,2,4,3)
\\
&\hspace{0.8cm}
+ \big( C^{NP}_{3412} + C^{NP}_{1234} \big) A_4^{[1]}(1,3,2,4)
\\
&\hspace{0.8cm}
+ 2 R^{P_1}_{1234} A_4^{[1/2]}(1,2,3,4) + 2 R^{P_1}_{1243} A_4^{[1/2]}(1,2,4,3)
\\
&\hspace{0.8cm}
+ \big( R^{NP_3}_{3412} + R^{NP_3}_{1234} \big) A_4^{[1/2]}(1,3,2,4)
\Big]\, + \, \mathcal{C}(234),
\end{split}
\label{unrendiv}
\end{align}
where we used the fact that $R^{P_1}_{1234}=R^{P_2}_{1234}$.
This matches eq.~\eqref{divpred} at the level of the
$(-s)^{-\epsilon}/\epsilon^2$ term, i.e.~except for the term
proportional to $b_0^{R_0}$.

Now the expression~\eqref{unrendiv} is for the unrenormalized two-loop
amplitude, whereas the Catani formula~\eqref{divpred} predicts the UV
renormalized one.  The renormalized amplitude
$\mathcal{A}_4^\text{2-loop,ren.}$ is given by adding the
$\overline{\text{MS}}$ counterterm
\be
\label{CT}
- 4 g^2 c_\Gamma b_0^{R_0}
 \frac{1}{\epsilon}\mathcal{A}_4^\text{1-loop}(1,2,3,4).
\ee
No other terms are needed due to the vanishing of the all-plus helicity
amplitude at tree level.

To arrive at the term proportional to $b_0^{R_0}$ in eq.~\eqref{divpred},
we use the color conservation identity 
\be
\sum_{i=1}^{n}\mathbf{T}_i=0
\ee
to write
\be
nC_A|\mathcal{M}_n\rangle = \sum_{i=1}^{n}\mathbf{T}_i^2|\mathcal{M}_n\rangle
= -2\sum_{1\leq i<j\leq n}\mathbf{T}_i\cdot \mathbf{T}_j|\mathcal{M}_n\rangle.
\ee
This identity allows us to write the counterterm~\eqref{CT}
in our notation as
\begin{align}
\begin{split}
-4g^2c_\Gamma b_0^{R_0}
\frac{1}{\epsilon}\mathcal{A}_4^\text{1-loop}(1,2,3,4)
&=-g^2c_\Gamma\frac{b_0^{R_0}}{C_A}\frac{1}{\epsilon}
\big(4C_A\mathcal{A}_4^\text{1-loop}(1,2,3,4)\big)
\\
&= -2g^2c_\Gamma\frac{b_0^{R_0}}{C_A}\frac{1}{\epsilon}
\sum_{1\leq i<j\leq 4}(if^{b_ia_ic})(if^{ca_jb_j})
\\
&\hspace{0.5cm}\times
A^\text{1-loop}_4(1^{a_1},\dotsc,i^{b_i},\dotsc,j^{b_j},\dotsc, 4^{a_4}).
\end{split}
\end{align}
Now it matches precisely the $b_0^{R_0}$-containing term of eq.~\eqref{divpred},
in the form of eqs.~\eqref{Aijsum} and \eqref{Aij}.

Thus, once the UV counterterm is included, we have exact agreement
between the infrared divergences of the renormalized two-loop amplitude
and the ones predicted by eq.~\eqref{divpred}.
In other words, the non-$b_0^{R_0}$, $1/\epsilon^2$ term
of eq.~\eqref{divpred} precisely matches the divergences of the
unrenormalized two-loop amplitude.

Next we evaluate eq.~\eqref{divpred}, but including also the ${\cal O}(\e^0)$
terms.  We subtract the result from the UV renormalized two-loop amplitude,
in order to obtain the Catani finite remainder,
$\mathcal{M}_4^{(2),\text{fin}}$.
This result exactly yields the CCA bootstrap formula~\eqref{KCformula}.
In other words, eq.~\eqref{KCformula} gives the IR-subtracted two-loop
amplitude. In the next section, we explore how to avoid an explicit
IR divergence and subtraction.

\section{Mass regularization}
\label{sec:massreg}

The requirement to subtract the IR divergences is unsatisfactory for the
following reasons. Firstly, the CCA bootstrap requires no such subtraction;
it is a completely finite procedure. Secondly, there is no dimensional
regularization prescription for sdYM, since its definition requires
the four-dimensional Levi-Civita tensor.
In some sense, the IR subtraction remedies a problem that is introduced
by our lack of understanding of how to regulate Feynman integrals in sdYM.
We remedy this by regulating the internal propagators that give rise to
IR divergences of the loop momenta with a particle mass. Then we can
take $\e\to0$ without encountering any poles in $\e$.

There is a fundamental difference between mass regularization and
dimensional regularization in when small terms can be neglected.
In dimensional regularization, divergences are powers of $1/\e$,
whose degree rises with the loop order.  Therefore terms suppressed
by powers of $\e$ in lower-loop amplitudes generally have to be retained.
On the other hand, when regulating with a particle mass $m$, divergences
are logarithmic in $m$. Hence power-suppressed contributions can
always be dropped, because any positive power of $m$ vanishes much faster
than (any power of) $\ln m$ increases, in the limit $m\to0$.

Mass regularization has previously been used in planar $\mathcal{N}=4$
supersymmetric
YM~\cite{Alday:2009zm,Henn:2010bk,Bourjaily:2013mma,Bourjaily:2019jrk,%
Arkani-Hamed:2023epq}.
Our method for assigning a mass to the propagators differs from these
examples.  Indeed, planar $\mathcal{N}=4$ supersymmetric YM
has dual conformal symmetry, which is closely related to these
regularization schemes.  It is not clear whether one can find a fully
consistent massive regulator of nonplanar YM theory, given that
the number of helicity states for massive vector bosons does not match
the massless case.
Hence, we do not claim that our method can consistently reproduce
the correct IR divergences in the massless limit for more complicated
amplitudes or integrals.  We are merely regulating the few divergent
integrals that appear in the all-plus four-point case.

In our scheme, a propagator is given a mass $m$ when a limit of
the loop momentum that puts it on shell also results in one or more
of its neighboring propagators going on shell.
The mass prevents the other propagators from diverging when the initial
one does.  For example, in the case of diagram (a) of
fig.~\ref{fig:integrals}, when the loop momentum $p\rightarrow k_1$,
all three of $(p-k_1)^2$, $p^2$, and $(p-k_1-k_2)^2$ approach zero.
So the mass $m$ is added to the first propagator
$(p-k_1)^2\mapsto(p-k_1)^2-m^2$.  Now, no two or more neighboring propagators
can simultaneously diverge.  We could have achieved the same result
by adding a mass $m$ to both the $p^2$ and $(p-k_1-k_2)^2$ propagators;
however, adding the mass to $(p-k_1)^2$ is the minimal solution and leads
to very simple integrals.

It is unnecessary to add a mass to propagators containing loop momentum
$\ell$ if the integral has a numerator factor of $(\lambda_\ell^2)^n$
for some positive integer $n$. This purely $(-2\eps)$-component of
$\ell$ vanishes when $\ell$ is purely four-dimensional, which in turn
prevents the appearance of the soft or collinear IR singularity associated
with the divergence of a propagator.
This argument includes cases where $\ell$ is a sum (or difference)
of loop momenta.

For the integrals appearing in eqs.~\eqref{gluprimamps} and \eqref{primamps}
and depicted in fig.~\ref{fig:integrals},
only the following replacements are necessary:
\begin{align}
\begin{split}
\mathcal{I}^P_4[\lambda_p^2\lambda_{p+q}^2]:\hspace{1cm}
(q-k_4)^2&\mapsto(q-k_4)^2-m^2,
\\
\mathcal{I}^P_4[\lambda_q^2\lambda_{p+q}^2]:\hspace{1cm}
(p-k_1)^2&\mapsto(p-k_1)^2-m^2,
\\
\mathcal{I}^{NP}_4[\lambda_p^2\lambda_{p+q}^2]:\hspace{1cm}
(q-k_2)^2&\mapsto(q-k_2)^2-m^2,
\\
\mathcal{I}^{NP}_4[\lambda_q^2\lambda_{p+q}^2]:\hspace{1cm}
(p-k_1)^2&\mapsto(p-k_1)^2-m^2,
\\
\mathcal{I}^{NP}_4[\lambda_p^2\lambda_q^2]:\hspace{0.35cm}
(p+q+k_3)^2&\mapsto(p+q+k_3)^2-m^2.
\end{split}
\end{align}
These new integrals can be evaluated directly using Feynman parameters, giving
\begin{align}
\label{massPint}
\mathcal{I}^P_{4,m^2}[\lambda_p^2\lambda_{p+q}^2]
=\mathcal{I}^P_{4,m^2}[\lambda_q^2\lambda_{p+q}^2]
=\mathcal{I}^{NP}_{4,m^2}[\lambda_p^2\lambda_q^2]
&=\frac{s^{-1}}{6(4\pi)^4}\big[\Li_2(1+s/m^2) - \zeta_2\big]
+ {\cal O}(\eps),
\\
\label{massNPint}
\mathcal{I}^{NP}_{4,m^2}[\lambda_p^2\lambda_{p+q}^2]
=\mathcal{I}^{NP}_{4,m^2}[\lambda_q^2\lambda_{p+q}^2]
&= {\cal O}(\eps),
\end{align}
where $\zeta_2=\zeta(2)=\Li_2(1)=\pi^2/6$.
The bow-tie integrals remain unchanged and are given in eq.~\eqref{bowties}.
In the $m\to0$ limit, eq.~\eqref{massPint} has the asymptotic behavior
\be
\frac{s^{-1}}{6(4\pi)^4}\big[\Li_2(1+s/m^2) - \zeta_2\big]
\sim-\frac{s^{-1}}{6(4\pi)^4}
\bigg[
\frac{1}{2}\ln^2\bigg(\frac{m^2}{-s}\bigg)+2\zeta_2
\bigg]\ +\ {\cal O}(m^2).
\ee
The divergent log term agrees with the leading-order divergent term in
the dimensionally-regulated integrals eqs.~\eqref{dimregPint}
and \eqref{dimregNPint}, i.e. the coefficient of the leading-order
$1/\eps^2$ equals the coefficient of $\tfrac{1}{2}\ln^2(-m^2/s)$.
Although it is necessary to take $m\to0$ to make apparent the
IR divergence, we will continue to work with the expression for
generic $m$, eq.~\eqref{massPint}, as it will not affect our analysis below.
	
These mass-regulated integrals are much simpler than their purely
dimensionally-regulated counterparts. We can understand these results
heuristically by considering the IR divergences appearing in the original
integrals. 
	
First consider eq.~\eqref{massPint}. When $m=0$, the divergent terms
in $\eps$ are given by eq.~\eqref{intdivpart}, the massless triangle
times the massless box. The mass-regulated planar integral in
eq.~\eqref{massPint} should factorize similarly when $m\to0$.
The mass-regulated triangle is
\be
\label{masstri}
\mathcal{I}^\text{1-loop}_{3,m^2}[1](s)
= \frac{i}{(4\pi)^2}s^{-1}\big[\Li_2(1+s/m^2) - \zeta_2\big]
+ {\cal O}(\eps),
\ee
and the box to zeroth order in $\eps$ is
\be
\mathcal{I}^\text{1-loop}_4[\lambda_\ell^4] 
= -\frac{i}{(4\pi)^2}\frac{1}{6}
+ {\cal O}(\eps).
\ee
So, eq.~\eqref{massPint} indeed agrees with the divergence statement
when $m\to0$. Surprisingly though, these mass-regulated
two-loop integrals are \textit{exactly} the product of the mass-regulated
triangle and the massless box, even for generic mass $m$.

As a check on the results, consider the $s$-channel cut in four dimensions
of $\mathcal{I}^P_{4,m^2}[\lambda_{p}^2\lambda_{p+q}^2]$, where we cut the
propagators neighboring the massive one, which corresponds to cutting
the right box vertically in fig.~\ref{fig:integrals}(a).
Indeed, the unitarity cuts can be performed in four dimensions since the
mass-regulated double box is finite \textit{a priori}. This produces a
factor of the massless box within the phase-space integral,
which is \textit{constant} in four dimensions. So, the massless box can be
factored out of the phase-space integral, and what remains within the
integral is nothing more than the $s$-channel cut of the mass-regulated
triangle. In other words,
\be
\label{discdbbox}
\text{Disc}_{s>0}~\mathcal{I}^P_{4,m^2}[\lambda_{p}^2\lambda_{p+q}^2]
= \mathcal{I}^\text{1-loop}_4[\lambda_p^4]~\text{Disc}_{s>0}~
\mathcal{I}^\text{1-loop}_{3,m^2}[1].
\ee
This is the only non-vanishing four-dimensional cut of the double-box,
since all other cuts vanish due to the vanishing of the $\lambda^2$
numerator factors in four dimensions. The discontinuity of the
mass-regulated triangle is easily computed from eq.~\eqref{masstri}
to be
\be
\text{Disc}_{s>0}~\mathcal{I}^\text{1-loop}_{3,m^2}[1]
= \frac{i}{(4\pi)^2}2\pi i \, \frac{\log(1+s/m^2)}{s} \,.
\ee
Since $\big|\mathcal{I}^\text{1-loop}_{3,m^2}[1](s)\big|\to0$ sufficiently fast
as $|s|\to\infty$, we can use a dispersion relation to compute the
triangle integral from its discontinuity along $s>0$. In other words,
\be
\mathcal{I}^P_{4,m^2}[\lambda_{p}^2\lambda_{p+q}^2](s)
=\frac{\mathcal{I}^\text{1-loop}_4[\lambda_p^4]}{2\pi i}
\int_{0}^{\infty}
dx~\frac{\text{Disc}_{x\geq0}~\mathcal{I}^\text{1-loop}_{3,m^2}[1]}{x-s}
=\mathcal{I}^\text{1-loop}_4[\lambda_p^4]
\, \mathcal{I}^\text{1-loop}_{3,m^2}[1](s).
\ee
	
We can understand the vanishing of the integrals in eq.~\eqref{massNPint}
by again understanding the IR divergences when $m=0$. Consider the
divergences of $\mathcal{I}^{NP}_{4,m^2}[\lambda_p^2\lambda_{p+q}^2]$.
The only propagators that are not suppressed by a $(-2\eps)$-component
of the loop momenta are $q^2$ and $(q-k_2)^2$. When the latter goes on shell,
the former does as well.  This gives a ${\cal O}(1/\eps)$ divergent term,
since they are neighboring propagators, multiplied by a box with a
doubled propagator, denoted by
$\mathcal{I}^\text{1-loop}_4[\lambda^4/(p-k_1)^2]$,
which is ${\cal O}(\eps)$ because it is related to an UV finite integral in
six dimensions.  The result is an integral that begins at ${\cal O}(\eps^0)$.
Following our procedure for mass regularizing, the only propagator given a
mass is $(q-k_2)^2$. This replaces the ${\cal O}(1/\eps)$
term by a ${\cal O}(\eps^0)$ one,
but it is still multiplied by a box of ${\cal O}(\eps)$.
Thus, the mass-regulated integral is ${\cal O}(\eps)$.  Notice also
that there is no four-dimensional unitarity cut of this integral,
so its vanishing is consistent with the vanishing of its cuts.

Mass regularization of the integrals renders the primitive
amplitudes~\eqref{gluprimamps} and \eqref{primamps} much simpler.
Substituting eqs.~\eqref{massPint} and \eqref{massNPint}
into eq.~\eqref{primamps} yields
\begin{align}
\label{primampsmass}
\begin{split}
A^{P_1}_{F1234} =&~
\frac{\rho}{(4\pi)^4}\bigg\{-\frac{1}{3}\big[\Li_2(1+s/m^2) - \zeta_2\big]
+\frac{1}{9}\bigg(\frac{t}{s}-4\bigg) + 1 \bigg\} + {\cal O}(\eps),
\\
A^{P_2}_{F1234} =&~
\frac{\rho}{(4\pi)^4}\bigg\{-\frac{1}{3}\big[\Li_2(1+s/m^2) - \zeta_2\big]
+\frac{1}{9}\bigg(\frac{t}{s}-4\bigg) + 1 \bigg\} + {\cal O}(\eps),
\\
A^{P_3}_{F1234} =&~ -\frac{\rho}{(4\pi)^4}\frac{1}{2},
\\
A^{P_4}_{F1234} =&~ -\frac{\rho}{(4\pi)^4}
\bigg\{
\frac{1}{9}\bigg(\frac{t}{s}-4\bigg) + \frac{1}{2}
\bigg\},
\\
A^{NP_1}_{F1234} =&~{\cal O}(\eps),
\\
A^{NP_2}_{F1234} =&~{\cal O}(\eps),
\\
A^{NP_3}_{F1234} =&~-\frac{\rho}{(4\pi)^4}
\frac{1}{3}\big[\Li_2(1+s/m^2) - \zeta_2\big] + {\cal O}(\eps).
\end{split}
\end{align}
The above primitive amplitudes are now a sum of rational terms and
terms of uniform transcendental weight two, which contain the mass
regulator $m$. In particular, $A^{NP_1}_F$ and $A^{NP_2}_F$ now vanish at
${\cal O}(\eps^0)$, and $A^{P_1}_F=A^{P_2}_F$ and $A^{NP_3}_F$ share the same
transcendental terms. Inspection of eqs.~\eqref{A411} and \eqref{A430}
shows that no transcendental terms remain in $A_{4;1;1}$ and $A_{4;3;0}$
when eq.~\eqref{primampsmass} is used,
\be
\label{A411funcmass}
A_{4;1;1}(1,2,3,4) = \frac{12\rho}{(4\pi)^4} + {\cal O}(\eps),
\ee
and
\be
\label{A430funcmass}
A_{4;3;0}(1,2,3,4) = \frac{24\rho}{(4\pi)^4} + {\cal O}(\eps),
\ee
and we now obtain complete agreement with eq.~\eqref{KCformula}.
Notice that this is true even without taking the limit $m\to0$. 

\section{Conclusions and outlook}
\label{sec:conclusions}

In this paper, we used previously-derived
results~\cite{Bern:2000dn,Bern:2002zk} and color algebra to perform
a check in dimensional regularization
of the two-loop four-point all-plus result~\eqref{KCformula}
from the CCA bootstrap.  The primitive amplitudes of
refs.~\cite{Bern:2000dn,Bern:2002zk} begin at ${\cal O}(1/\eps^2)$ and
contain transcendental functions.  By placing the matter in the
representation $R_0$, we found that the color-ordered two-loop amplitudes
in this theory contain both $1/\eps$ poles and transcendental terms.
At first sight, this might seem to contradict eq.~\eqref{KCformula}.
However, Catani's universal factorization formula \eqref{Catani} exactly
predicts these terms, and our computation agrees with \eqref{KCformula}
after we subtract the universal IR divergences.

The discrepancy arises due to the non-vanishing of the one-loop amplitude
in this theory at ${\cal O}(\eps)$. We remedy this by mass-regulating the
already dimensionally-regulated integrals comprising the primitive
amplitudes \eqref{gluprimamps} and \eqref{primamps}.
All appearances of the dimension regulator $\eps$ are replaced by
expressions involving the mass regulator $m$, resulting in finite quantities
when $m\neq0$. The new mass-regulated amplitudes give exact agreement with
eq.~\eqref{KC4pt}, even for generic mass $m$.
Removing the dependence on $\eps$ is essential for comparing the YM and
sdYM results. The self-dual equations explicitly depend on the Levi-Civita
tensor, which does not have a sensible definition for non-integer dimensions.
So, the CCA bootstrap must implicitly use a different IR regularization scheme
that involves keeping all momenta in four dimensions. Mass regularization
appears to be such a scheme, at least at four points, and a suitable
mass regularization for higher-point all-plus amplitudes seems likely
to lead to agreement with eq.~\eqref{KCnpt} as well.
	
Despite the discrepancy between the sdYM form factor and the YM amplitude
in dimensional regularization, we are confident in the validity of
eq.~\eqref{KCnpt} when using a suitable mass regulator. Taking all possible
four-dimensional unitarity cuts of the two-loop all-plus sdYM form factor
shows that it cannot have any branch cuts in the $R_0$ theory.
Moreover, the vanishing of the one-loop form factor forces the two-loop
one to behave like a tree-level form factor in collinear limits,
suggesting that the two-loop one is finite.
We believe that eq.~\eqref{KCnpt} predicts the finite remainder of the
YM amplitude in dimensional regularization. In particular, we predict that
\be
\label{twoloopnpt}
\mathcal{A}_n^\text{2-loop} = \mathcal{A}_n^\text{2-loop}\big|_\text{pred. div.}
+ \mathcal{A}_{n,\text{sdYM}}^\text{2-loop},
\ee 
where $\mathcal{A}_n^\text{2-loop}\big|_\text{pred. div.}$ is the predicted
IR divergence of Catani given by eqs.~\eqref{Aijsum} and \eqref{Aij},
including its ${\cal O}(\e^0)$ parts,
and $\mathcal{A}_{n,\text{sdYM}}^\text{2-loop}$ is the two-loop result
computed from the CCA bootstrap given by eq.~\eqref{KCnpt}.
Evaluating $\mathcal{A}_n^\text{2-loop}\big|_\text{pred. div.}$
to ${\cal O}(\e^0)$requires
knowing the one-loop all-plus $n$-point amplitude to $\mathcal{O}(\eps^2)$.
A closed-form formula is not known for this, but in principle it can be
computed for each $n$ by using a basis of scalar integrals.
The complete basis to all-orders in $\eps$ includes pentagons, boxes,
bubbles and triangles~\cite{%
Bern:1992em,Bern:1993kr,Giele:2008ve,Ellis:2008ir,Badger:2008cm},
the coefficients of which can be computed using $D$-dimensional
unitarity~\cite{Bern:1996ja,Britto:2020crg}.

The combination of the single-trace term computed in
ref.~\cite{Costello:2023vyy} and the double-trace term that we have computed,
eq.~\eqref{KCnpt}, is a complete two-loop $n$-point result for
a non-supersymmetric gauge theory with matter. We have conjectured that
the YM amplitude has the form given by \eqref{twoloopnpt} with dimensional
regularization as the IR regulator. Mass regularization of the four-point
integrals allows for complete agreement between the YM amplitudes and the
sdYM form factor.  We further conjecture that this scheme, and perhaps
other IR regularization schemes which do not change the dimensions of
spacetime, give complete agreement between the two approaches at higher points.
In light of the simple behavior of the two-loop all-plus four-point
amplitudes when dimensional regularization is combined with suitable
mass regularization, it may be worth investigating similar mass regularization
for other types of gauge theory amplitudes.
Although the two-loop all-plus $n$-point
amplitude was not computed in QCD, where the fermions are in
the representation $N_F\,(F\oplus\bar{F})$,
methods similar to the CCA bootstrap, perhaps when combined with other
bootstrap methods, may lead to analytic progress in the computation of
higher-order corrections in more realistic theories.

\acknowledgments

We thank Kevin Costello, Roland Bittleston, Song He, Henrik Johansson,
and Oliver Schlotterer for helpful discussions, and Kevin Costello
for useful comments on the draft.  We are particularly grateful
to Kevin Costello for hosting us at the Perimeter Institute for Theoretical
Physics (PI), where part of this work was completed.
This research was supported by the US Department of Energy under contracts
DE--AC02--76SF00515 and DE--FOA--0002705, KA/OR55/22 (AIHEP),
and was also supported in part by PI.
Research at PI is supported by the Government of Canada through the
Department of Innovation, Science and Economic Development and by the
Province of Ontario through the Ministry of Colleges and Universities.

\appendix

\section{Colorful identities}
\label{sec:colorid}

In order to evaluate the color factors in terms of traces over the
fundamental representation
without any contracted indices, we make use of various
$SU(N)$ identities.  In this appendix, we let $R$ be an arbitrary
irreducible representation of $SU(N)$. The fundamental (defining)
representation is denoted by $F$, and $G$ denotes the adjoint
representation.

In the main text, we normalize the generators such that the Dynkin index
of the fundamental representation $T_F$ is unity, i.e.
\be
  \tr(t^at^b)=\delta^{ab}\quad \iff\quad T_F=1.
\ee
Furthermore, we define the $SU(N)$ structure constants $f^{abc}$ to
be real and normalized such that
\be
  i f^{abc} = \tr([t^a,t^b]t^c).
\ee

We will make use of color diagrams to describe one- and two-loop
color factors. The rules for evaluating color diagrams are
\begin{align}
	\label{diagramrules}
	\begin{split}
		\begin{tikzpicture}[baseline={(0,0)},scale=0.75]
			\begin{feynman}
				\vertex (g1) at (0,1.5) {\(a\)};
				\vertex (g2) at (1.5*1.7321/2,-1.5/2) {\(b\)};
				\vertex (g3) at (-1.5*1.7321/2,-1.5/2) {\(c\)};
				\vertex (v1) at (0,0);
				\diagram*{
					(g1) -- [gluon] (v1),
					(g2) -- [gluon] (v1),
					(g3) -- [gluon] (v1),
				};
			\end{feynman}
		\end{tikzpicture}
		&= if^{abc}
		\\
		\begin{tikzpicture}[baseline={(0,0)},scale=0.75]
			\begin{feynman}
				\vertex (g1) at (-1.5,0) {\(a\)};
				\vertex (g2) at (1,0) {\(b\)};
				\diagram*{
					(g1) -- [gluon] (g2)
				};
			\end{feynman}
		\end{tikzpicture}
		&= \delta^{ab}
		\\
		\begin{tikzpicture}[baseline={(0,0)},scale=0.75]
			\begin{feynman}
				\vertex (ti) at (-1.5*1.7321/2,0) {\(i\)};
				\vertex (tj) at (1.5*1.7321/2,0) {\(j\)};
				\vertex (v) at (0,0);
				\vertex (g) at (0,1.5) {\(a\)};
				\diagram*{
					(ti) -- [fermion, edge label=\(R\)] (v) 
					-- [fermion, edge label=\(R\)] (tj);
					(g) -- [gluon] (v)
				};
			\end{feynman}
		\end{tikzpicture}
		&= (t^a_R)^i_{~j}
		\\
		\begin{tikzpicture}[baseline={(0,0)},scale=0.75]
			\begin{feynman}
				\vertex (ti) at (-1.5,0) {\(i\)};
				\vertex (tj) at (1,0) {\(j\)};
				\diagram*{
					(ti) -- [fermion, edge label=\(R\)] (tj)
				};
			\end{feynman}
		\end{tikzpicture}
		&= (\delta_R)^i_{~j}
	\end{split}	
\end{align}
where $t^a_R$ is an $SU(N)$ generator in a representation $R$.
If the ``$R$'' is omitted, then it is implied that the generators are
in the fundamental/anti-fundamental representation. The graphical
depiction of the antisymmetric tensor product of the fundamental
$\asym$ is given in fig.~\ref{fig:asym}.

In this appendix, we will keep the Dynkin index of the fundamental
representation $T_F$ arbitrary. It is set to 1 outside this appendix.

Recall that the quadratic Casimir in $R$ is defined by
\be
  t_R^at_R^a=C_R\cdot\text{id}_R \,.
\ee
Two other contractions of generators that appear in the computations are
\begin{align}
  t_R^ct_R^at_R^c &= \bigg(C_R-\frac{C_G}{2}\bigg)t_R^a \,,
\\
t_R^ct_R^at_R^bt_R^c &= (C_R-C_G)t_R^at_R^b + (if^{adc})(if^{ceb})t_R^dt_R^e \,.
\end{align}
The $SU(N)$ Fierz identity,
\be
 (t^a)^i_j(t^a)^k_l
 = T_F\delta^i_l\delta^k_j - \frac{T_F}{N}\delta^i_j\delta^k_l \,,
\ee
when in the presence of other matrices and inside traces is given by
\begin{align}
 \tr(Xt^aYt^aZ) &= T_F\tr(Y)\tr(XZ) - \frac{T_F}{N}\tr(XYZ),
\\
 \tr(Wt^aX)\tr(Yt^aZ) &= T_F\tr(ZYXW) - \frac{T_F}{N}\tr(WX)\tr(YZ),
\\
 \tr(Wt^aX)Yt^aZ &= T_FYXWZ-\frac{T_F}{N}\tr(WX)YZ.
\end{align}

The trace over the exterior square of the fundamental $\asym$ has a
realization as traces over $F$. Letting $P:F\otimes F\to F\otimes F$
be the permutation operator, the exterior square is the image of the
projector $\tfrac{1}{2}(1-P)$. Thus, the trace over $\asym$ is given by
\be
 \tr_\asym(t^{a_1}\cdots t^{a_n}) 
= \tr_{F\otimes F}(t^{a_1}\cdots t^{a_n}\tfrac{1}{2}(1-P)).
\ee
The generators of $F\otimes F$ are related to the generators of $F$ by
\be
\label{FFtensor}
  t^a_{F\otimes F} = t^a\otimes 1 + 1\otimes t^a,
\ee
which implies that there are $2^n$ contributions to the trace
over $n$ generators $t^{a_i}_{F\otimes F}$, according to the choice
of first or second term in eq.~\eqref{FFtensor}.
In other words,
\be
  \tr_{F\otimes F}(t^{a_1}\cdots t^{a_n})
 = \sum_{I\subset(1,\dotsc,n)}\tr(t_I)\tr(t_{I^c}),
\ee
where $t_I$ denotes the product $t^{a_{i_1}}\cdots t^{a_{i_m}}$ for
$I=(i_1,\dotsc,i_m)$ with the ordering inherited from the
ordered list $(1,\dotsc,n)$, and $I^c$ is the complement of $I$,
again with the inherited ordering. Similarly,
\be
\tr_{F\otimes F}(t^{a_1}\cdots t^{a_n}P) =\sum_{I\subset(1,\dotsc,n)}\tr(t_It_{I^c}),
\ee
so that
\be
  \tr_\asym(t^{a_1}\cdots t^{a_n}) = \frac{1}{2}\sum_{I\subset(1,\dotsc,n)}
		\big[\tr(t_I)\tr(t_{I^c}) - \tr(t_It_{I^c})\big].		
\ee
Note that 
\be
  t_\emptyset = \mathbf{1},
\ee
with this notation, so
\be
 \tr(t_\emptyset)=\tr(\mathbf{1})=N.
\ee
The associated color diagram for the trace over $\asym$ is shown in
fig.~\ref{fig:antisym}. It uses the diagramatic rules for
antisymmetrizing two lines shown in fig.~\ref{fig:asym}.
	
\begin{figure}
\centering
\be
\nonumber
\begin{tikzpicture}[baseline={(0,0)},scale=0.75]
		\begin{feynman}
			\vertex (g1) at (0,-2);
			\vertex (g2) at (0,2);
			\vertex (g3) at (4,2);
			\vertex (g4) at (4,-2);
			\vertex (v1) at (1,-1);
			\vertex (v2) at (1,1);
			\vertex (v3) at (3,1);
			\vertex (v4) at (3,-1);
			\vertex (w1) at (1.25,-0.75);
			\vertex (w2) at (1.25,0.75);
			\vertex (w3) at (2.75,0.75);
			\vertex (w4) at (2.75,-0.75);
			\vertex (b1) at (1.65,-1.75);
			\vertex (b2) at (1.65,-0.75);
			\vertex (b3) at (2.35,-0.75);
			\vertex (b4) at (2.35,-1.75);
			\diagram*{
				(g1) -- [gluon] (v1),
				(v2) -- [gluon] (g2),
				(g3) -- [gluon] (v3),
				(v4) -- [gluon] (g4),
				(v1) -- [quarter left, fermion] (v2) 
				     -- [quarter left, fermion] (v3) 
				     -- [quarter left, fermion] (v4) 
				     -- [quarter left, fermion] (v1),
				(w1) -- [quarter left, fermion] (w2) 
				     -- [quarter left, fermion] (w3) 
				     -- [quarter left, fermion] (w4) 
				     -- [quarter left, fermion] (w1),
				(b1) -- (b2) -- (b3) -- (b4) -- (b1),
			};
		\end{feynman}
	\end{tikzpicture}
	=
	\frac{1}{2}
	\begin{tikzpicture}[baseline={(0,0)},scale=0.75]
		\begin{feynman}
			\vertex (g1) at (0,-2);
			\vertex (g2) at (0,2);
			\vertex (g3) at (4,2);
			\vertex (g4) at (4,-2);
			\vertex (v1) at (1,-1);
			\vertex (v2) at (1,1);
			\vertex (v3) at (3,1);
			\vertex (v4) at (3,-1);
			\vertex (w1) at (1.25,-0.75);
			\vertex (w2) at (1.25,0.75);
			\vertex (w3) at (2.75,0.75);
			\vertex (w4) at (2.75,-0.75);
			\diagram*{
				(g1) -- [gluon] (v1),
				(v2) -- [gluon] (g2),
				(g3) -- [gluon] (v3),
				(v4) -- [gluon] (g4),
				(v1) -- [quarter left, fermion] (v2) 
				-- [quarter left, fermion] (v3) 
				-- [quarter left, fermion] (v4) 
				-- [quarter left, fermion] (v1),
				(w1) -- [quarter left, fermion] (w2) 
				-- [quarter left, fermion] (w3) 
				-- [quarter left, fermion] (w4) 
				-- [quarter left, fermion] (w1),
			};
		\end{feynman}
	\end{tikzpicture}
	- \frac{1}{2}
	\begin{tikzpicture}[baseline={(0,0)},scale=0.75]
		\begin{feynman}
			\vertex (g1) at (0,-2);
			\vertex (g2) at (0,2);
			\vertex (g3) at (4,2);
			\vertex (g4) at (4,-2);
			\vertex (v1) at (1,-1);
			\vertex (v2) at (1,1);
			\vertex (v3) at (3,1);
			\vertex (v4) at (3,-1);
			\vertex (w1) at (1.25,-0.75);
			\vertex (w2) at (1.25,0.75);
			\vertex (w3) at (2.75,0.75);
			\vertex (w4) at (2.75,-0.75);
			\vertex (b1) at (2+1.4*0.34202,-1.4*0.93969);
			\vertex (b2) at (2-1.4*0.34202,-1.4*0.93969);
			\vertex (b3) at (2.1,-1.049161396);
			\vertex (b4) at (1.9,-1.141211056);
			\diagram*{
				(g1) -- [gluon] (v1),
				(v2) -- [gluon] (g2),
				(g3) -- [gluon] (v3),
				(v4) -- [gluon] (g4),
				(v1) -- [quarter left, fermion] (v2) 
				-- [quarter left, fermion] (v3) 
				-- [quarter left, fermion] (v4),
				(w1) -- [quarter left, fermion] (w2) 
				-- [quarter left, fermion] (w3) 
				-- [quarter left, fermion] (w4),
				(b1) -- (w1),
				(w4) -- (b3),
				(b4) -- (b2),
			};
			\draw ([shift=(-20:1.4)] 2, 0) arc (-20:-70:1.4);
			\draw ([shift=(200:1.4)] 2, 0) arc (200:250:1.4);
\end{feynman}
\end{tikzpicture}
\ee
\caption{Color diagram for the trace over the representation $\asym$
in terms of traces over the fundamental $F$.}
\label{fig:antisym}
\end{figure}
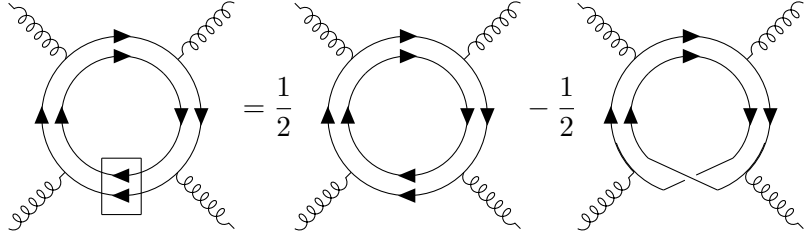
		
The Dynkin index of $\asym$ is read off from
\be
\tr_\asym(t^at^b)=\frac{1}{2}\big[2N\tr(t^at^b)-4\tr(t^at^b)\big]
= T_F(N-2)\delta^{ab},
\ee
i.e.
\be
T_\asym = T_F(N-2).
\label{Tasym}
\ee
The quadratic Casimir is then
\be
C_\asym = \frac{T_\asym\dim G}{\dim\asym}
= 2T_F(N-1-2N^{-1}).
\ee

\section{Computation of the double-trace kinematic terms in eq.~\eqref{KCnpt}}
\label{sec:dtcomputation}
	
The color factors computed in ref.~\cite{Costello:2023vyy} leave out the
double-trace terms. Here, we recompute these color factors while
keeping track of the double-trace structure. Afterwards, we use the
CCA bootstrap to prove eq.~\eqref{dtterm}. In order to do this,
we first introduce some notation from ref.~\cite{Costello:2023vyy}.
	
The momenta for massless states satisfy
\be
p_{\alpha\dot{\alpha}} = \lam_\alpha\tilde{\lam}_{\dot{\alpha}},
\ee
where $\lam,\tilde{\lam}$ are two-component Weyl spinors.
We can scale $\lam,\tilde{\lam}$ while keeping the momentum $p$ fixed
such that
\be
\lam=(1,z).
\ee
The parameter $z$ is then the coordinate on the $\mathbb{CP}^1$ where
the chiral algebra lives. A massless state of energy $\omega$ is
described by a function of $z$ and $\tilde{\lam}$. For a set of $n$
outgoing momenta $\{p_i\}$, the familiar spinor brackets are defined by
\begin{align}
\label{spadef}
\spa{ij}&=2\pi i(z_i-z_j),
\\
\spb{ij}&=-\eps_{\dot{\alpha}\dot{\beta}}\tilde{\lam}_i^{\dot{\alpha}}
\tilde{\lam}_j^{\dot{\beta}}.
\end{align}

Positive- and negative-helicity states of a gauge theory are denoted by
\be
J^a[\omega\tilde{\lam}](z)~\text{and}~\tilde{J}^a[\omega\tilde{\lam}](z),
\ee
respectively, where $a$ is the color index. The states can be expanded in
a series in $\omega$ as
\begin{align}
\begin{split}
J^a[\omega\tilde{\lam}](z) &= \sum_{k}\omega^k J^a[k](z),
\\
\tilde{J}^a[\omega\tilde{\lam}](z) &= \sum_{k}\omega^k \tilde{J}^a[k](z),
\end{split}
\end{align}
where $J^a[k],\tilde{J}^a[k]$ are homogeneous polynomials of order $k$ in
$\tilde{\lam}$. These quantities are expanded further as
\begin{align}
\begin{split}
J^a[k](z)&=\sum_{r+s=k}\frac{1}{r!s!}
\big(\tilde{\lam}_{\dot{1}}\big)^r\big(\tilde{\lam}_{\dot{2}}\big)^s
J^a[r,s](z),
\\
\tilde{J}^a[k](z)&=\sum_{r+s=k}\frac{1}{r!s!}
\big(\tilde{\lam}_{\dot{1}}\big)^r\big(\tilde{\lam}_{\dot{2}}\big)^s
\tilde{J}^a[r,s](z).
\end{split}
\end{align}
The states $J^a[r,s],\tilde{J}^a[r,s]$ generate the (extended) chiral algebra
for pure sdYM living on the $z$-plane. These states should be thought
of as soft modes, since they result from an expansion in $\omega$. 

The OPEs in the chiral algebra correspond to collinear limits of states
in sdYM. At tree-level, the OPEs are
\begin{align}
J^a[\tilde{\lam}_i](z_i)J^b[\tilde{\lam}_j](z_j)&\sim
if^{abc}\frac{1}{\spa{ij}}J^c[\tilde{\lam}_i+\tilde{\lam}_j](z_i),
\\
J^a[\tilde{\lam}_i](z_i)\tilde{J}^b[\tilde{\lam}_j](z_j)&\sim
if^{abc}\frac{1}{\spa{ij}}\tilde{J}^c[\tilde{\lam}_i+\tilde{\lam}_j](z_i).
\end{align}
We have redefined the normalization of the $\tilde{\lam}$
in order to remove the appearance of the energy $\omega$.
Notice that the structure constants used here differ from that of
ref.~\cite{Costello:2023vyy} by a factor of $i$. The higher loop-order OPEs
(including those with matter) are found in ref.~\cite{Costello:2023vyy},
but they are not necessary for our purposes here.

As mentioned in the main body of the paper, correlation functions of the
chiral algebra in a given conformal block are form factors of sdYM with
an operator insertion $\mathcal{O}$ at a point in spacetime
corresponding to the
conformal block. We denote these correlators as
\be
\langle\mathcal{O}|J[\tilde{\lam}_1](z_1)\cdots
\tilde{J}[\tilde{\lam}_k](z_k)\cdots\rangle.
\ee
Expanding the external states as a sum of soft modes, we are left with
computing correlators of the form
\be
\langle\mathcal{O}|J[k_1](z_1)\cdots \tilde{J}[k_2](z_k)\cdots\rangle.
\ee
Since correlators on twistor space must not scale with dilations of
$\mathbb{R}^4$, the scaling dimensions of the external states must sum
to minus the scaling dimension of the operator. Positive-helicity states
$J[k]$ contribute dimension $-k$, while negative-helicity states
$\tilde{J}[k]$ contribute $-k-2$.

The OPEs constrain the poles of the correlators. In order to compute the
two-loop amplitude eq.~\eqref{KCnpt}, only knowledge of tree-level and
one-loop OPEs are needed. In particular, poles that involve $J[0]$ and
$J[1]$ insertions are dictated by tree-level and one-loop OPEs, respectively.

The chiral algebra also places constraints on terms which are regular in a
given limit. The algebra can be derived directly via Koszul
duality~\cite{Costello:2022wso}. This involves coupling $J[k]$ and
$\tilde{J}[k]$ to the gauge field and the auxiliary field of sdYM on
twistor space, which requires $J[k]$ to have a zero of order $2-k$ at
$z=\infty$ and $\tilde{J}[k]$ to have a pole of order $2+k$ at $z=\infty$
in order for the coupling to be well-defined.

Form factors of sdYM with the operator
${\cal O} = \tfrac{1}{2}\tr(B\wedge B)$
inserted at the origin give YM amplitudes when the sum of
the gluon momenta vanishes.
However, at two loops, the operator is chosen to be
\be
\label{2loopop}
\tfrac{1}{2}\tr(B\wedge B) + \hbar^2 C \tr(F\wedge F),
\ee
where $C$ is some constant. The $\tr(F\wedge F)$ term is added as a two-loop
counterterm, with $\hbar^2$ to remind us that the term is added at two loops.
This term is added in order to remove an all-plus-helicity two-loop
two-point correlator that can only be determined up to an overall constant $C$;
this addition also forces the two-loop three-point correlators to vanish. 
The operator $\tr(F\wedge F)$ is a total derivative,
which means that form factors with this operator vanish
when we impose that the momenta of the gluons add up to zero,
which we do when we pass to a scattering amplitude.  So in practice
we can neglect the second term in eq.~\eqref{2loopop}.

Since $\mathcal{O}$ has dimension four, and $J[1]$ dimension $-1$,
we consider the scale-invariant four-point correlator
\be
\langle
\mathcal{O}|J^{a_1}[1](z_1)J^{a_2}[1](z_2)J^{a_3}[1](z_3)J^{a_4}[1](z_4)
\rangle,
\label{fourptcorr}
\ee
where $\mathcal{O}$ means eq.~\eqref{2loopop} from now on.
Eq.~\eqref{fourptcorr} is determined by one-loop OPEs between any two
$J[1]$'s; hence we get a two-loop result.
It evaluates to (see ref.~\cite{Costello:2023vyy} for how this is computed)
\begin{align}
\label{4ptcorr}
\langle
\mathcal{O}|&J^{a_1}[1](z_1)J^{a_2}[1](z_2)J^{a_3}[1](z_3)J^{a_4}[1](z_4)\rangle
\nn\\
&= \frac{i}{(4\pi)^4} \frac{\spb{12}\spb{34}}{\spa{12}\spa{34}}
\frac{R^{a_1a_2a_3a_4}}{4}
\nn\\
&\hskip0.5cm - \frac{2i}{(4\pi)^4}
\frac{\spb{12}\spb{34}}{\spa{12}\spa{34}}
\frac{\spa{13}\spa{24}+\spa{14}\spa{23}}{\spa{12}\spa{34}}
\big( \tr(1234)+\tr(1432)-\tr(1243)-\tr(1342) \big)
\nn\\
&\hskip0.5cm + (1324) + (1423),
\end{align}
where the last line adds two more permutations,
and the color factor $R^{a_1a_2a_3a_4}$ is given by
\begin{align}
\label{Rcolorfactor}
\begin{split}
R^{a_1a_2a_3a_4} =&~ 4\Big(t_G^{(a_1}t_G^{a_2)}\Big)_{b_1b_2}
\Big(t_G^{(a_3}t_G^{a_4)}\Big)_{b_3b_4}
\Big(
-2\tr\big((b_1b_2)(b_3b_4)\big) + \tr(b_1b_3b_2b_4) + \tr(b_1b_4b_2b_3)
\Big)
\\
&+4\Big(t_G^{(a_1}t_G^{a_2)}\Big)_{b_1b_2}\tr_{R_0}\big((a_3a_4)(b_1b_2)\big)
+4\Big(t_G^{(a_3}t_G^{a_4)}\Big)_{b_1b_2}\tr_{R_0}\big((a_1a_2)(b_1b_2)\big)
\\
&-4\tr_{R_0}\big(c(a_1a_2)c(a_3a_4)\big).
\end{split}
\end{align}
The parentheses around color indices means to symmetrize on said indices.
Recall that $t_G^a$
are the generators of $SU(N)$ in the adjoint representation defined by
\be
(t_G^a)_{bc}=-if^{abc}.
\ee
Writing eq.~\eqref{Rcolorfactor} as traces over the fundamental without
contracted indices requires the use of the identities in
Appendix \ref{sec:colorid}. Doing so results in
\begin{align}
\begin{split}
R^{a_1a_2a_3a_4} =&~(24N-16-32N^{-1})
\big(
\tr(1234) + \tr(1243) + \tr(1342) + \tr(1432)
\big)
\\
&-(16+32N^{-1})
\big(
\tr(1324) + \tr(1423)
\big)
\\
&+(32+32N^{-1})
\big(
\tr(12)\tr(34) + \tr(13)\tr(24) + \tr(14)\tr(23)
\big) \,.
\end{split}
\end{align}
With this formula, eq.~\eqref{4ptcorr} can now be expressed as a sum over
permutations of different trace structures, resulting in
eqs.~\eqref{KC4pt}--\eqref{KC_A43}.

We now prove the formula~\eqref{dtterm} for the $n$-point double-trace
term by induction.  Eq.~\eqref{dtterm} clearly reproduces the $n=4$ case.
For the $n>4$ case, the correlator giving rise to
$A_{n;c}^\text{2-loop}(i_1,i_2,i_3,i_4)$ is
\be
\label{nptcorr}
\langle\mathcal{O}|\cdots J^{a_{i_1}}[1](z_{i_1})\cdots
J^{a_{i_2}}[1](z_{i_2})\cdots J^{a_{i_3}}[1](z_{i_3})\cdots 
J^{a_{i_4}}[1](z_{i_4})\cdots\rangle,
\ee
where ellipses indicate $J[0]$ insertions. Assume the $n$-th insertion is
a $J[0]$. Viewing eq.~\eqref{nptcorr} as a function of $z_n$, the poles
with respect to $z_n$ are dictated by the OPEs of $J^{a_n}[0](z_n)$ with
the other insertions. The OPEs are
\begin{align}
J^{a_m}[0](z_m)J^{a_n}[0](z_n)&\sim if^{a_ma_nb}\frac{1}{\spa{mn}}J^{b}[0](z_m),
\\
J^{a_m}[1](z_m)J^{a_n}[0](z_n)&\sim if^{a_ma_nb}\frac{1}{\spa{mn}}J^{b}[1](z_m).
\end{align}
The OPEs dictate that the residues at the simple poles $\spa{mn}$
will be $(n-1)$-point correlators.

The double-trace structures in eq.~\eqref{nptcorr} for $(n-1)$ points with
the ordering $1,2,\dotsc,n-1$ are
\be
\sum_{c=3}^{n-2}A_{n-1;c}^\text{2-loop}(i_1,i_2,i_3,i_4)
\tr(1\cdots c-1)\tr(c\cdots n-1).
\ee
Since we are only concerned with the ordering $1,2,\dotsc,n$ in the trace
structures for $n$ points, to determine the dependence on $z_n$
we only need to consider two OPEs, where the
point $z_n$ is near $z_c$ and where it is near $z_{n-1}$,
for a given $c\in\{3,\dotsc, n-2\}$.
Then the double-trace structure of eq.~\eqref{nptcorr} at a given $c$ has
the form
\begin{align}
\begin{split}
i&f^{a_{n-1}a_nb}\frac{1}{\spa{n-1,n}}
A_{n-1;c}^\text{2-loop}(i_1,i_2,i_3,i_4)
\tr(1\cdots c-1)\tr(c\cdots (n-2)b)
\\
&+if^{a_na_cb}\frac{1}{\spa{nc}}
A_{n-1;c}^\text{2-loop}(i_1,i_2,i_3,i_4)
\tr(1\cdots c-1)\tr(b(c+1)\cdots n-1)
\\
&= A_{n-1;c}^\text{2-loop}(i_1,i_2,i_3,i_4)
\bigg(
\frac{1}{\spa{n-1,n}}+\frac{1}{\spa{nc}}
\bigg)
\tr(1\cdots c-1)\tr(c\cdots n)
\\
&= A_{n-1;c}^\text{2-loop}(i_1,i_2,i_3,i_4)
\frac{\spa{n-1,c}}{\spa{n-1,n}\spa{nc}}
\tr(1\cdots c-1)\tr(c\cdots n)
\\
&= A_{n;c}^\text{2-loop}(i_1,i_2,i_3,i_4)\tr(1\cdots c-1)\tr(c\cdots n).
\end{split}
\end{align}
In the first equality, we performed the contractions between the structure
constants and the generators within the traces and kept only the
double-trace terms with the ordering $1,2,\dotsc,n$. In the second equality,
we used the definition of the angle spinor brackets in terms of the
$z_i$ variables, eq.~\eqref{spadef}. The last equality follows
by induction from the definition~\eqref{dtterm}.

Summing over $c$ then yields
\begin{align}
&\sum_{c=3}^{n-2}A_{n;c}^\text{2-loop}(i_1,i_2,i_3,i_4)
\tr(1\cdots c-1)\tr(c\cdots n)
\nn\\
&= \, \sum_{c=3}^{n-1}A_{n;c}^\text{2-loop}(i_1,i_2,i_3,i_4)
\tr(1\cdots c-1)\tr(c\cdots n),
\label{zninsertionfinal}
\end{align}
where we added $0=A_{n;n-1}^\text{2-loop}(i_1,i_2,i_3,i_4)$ in the second line,
which agrees with eq.~\eqref{dtterm}, since the condition
$c = n-1 \leq i_3 < i_4 \leq n$ is incompatible with $i_4 < n$, which
holds because a $J[0]$ is inserted at the $n$-th position.

In the above, we ignored the terms regular in $\spa{n-1,n}$ and $\spa{nc}$;
however, these terms are null since they would not allow $J^{a_n}[0](z_n)$
to have a second-order zero at $z_n=\infty$.  Eq.~\eqref{zninsertionfinal}
shows that the dependence on $z_n$ is compatible inductively
with eq.~\eqref{dtterm}.

Next we consider the dependence on $z_m$, when there is a $J[0]$ inserted
into eq.~\eqref{nptcorr} at $z_m$ for $m<n$.
The computation goes very similarly for the three cases:
$1\leq m< i_1$, $i_1<m<i_2$, and $i_3<m<i_4$.
The vanishing conditions for eq.~\eqref{dtterm} mean that the
$J^{a_m}[0]$ contributes to only one of the two traces in the double-trace
structure when only considering the ordering $1,2,\dotsc, n$.
The case $i_2<m<i_3$ differs slightly.
Taking $m=i_2+1$, the OPEs involving $J^{a_{i_2+1}}[0](z_{i_2+1})$ dictate
that there are simple poles at $z_{i_2+1}=z_j$ for
$j\in\{1,\dotsc n\}\setminus\{i_2+1\}$, and their residues are
$(n-1)$-point correlators with $J^{a_{i_2+1}}[0](z_{i_2+1})$ removed. 

The double-trace structure in the ordering $1,2,\dotsc,n$ of the
$(n-1)$-point correlator with this operator removed is
\be
\sum_{j=3}^{n-2}A_{n;c_j}^\text{2-loop}(i_1,i_2,i_3,i_4)
\tr(c_1\cdots c_{j-1})\tr(c_j\cdots c_{n-1}),
\ee
where $c_j$ is the $j$-th element of the ordered list
$(1,\dotsc,i_2,i_2+2,\dotsc,n)$. We can ignore terms with $j<i_2+1$, since
$A_{n-1;c_j}^\text{2-loop}(i_1,i_2,i_3,i_4)$ vanishes with this condition.
For $j>i_2+1$, the insertion $J^{a_{i_2+1}}[0](z_{i_2+1})$ only contributes
to the right trace in the double-trace structure. So the computation is
very similar to the $z_n$ case described above.

For $j=i_2+1$, $J^{a_{i_2+1}}[0](z_{i_2+1})$ must contribute
to both traces in the double-trace structure, since the generator
$t^{a_{i_2+1}}$ can be inserted in either of the traces in
\be
\tr(1\cdots i_2)\tr(i_2+2\cdots n)
\ee
while preserving the ordering $1,\dotsc,n$. 
Correspondingly, there are four poles instead of two, 
at $z_{i_2+1}\in\{z_1,z_{i_2},z_{i_2+2},z_n\}$, with residues
dictated by the OPEs:
\begin{align}
\begin{split}
i&f^{a_{i_2}a_{i_2+1}b}\frac{1}{\spa{i_2,i_2+1}}
A_{n-1;c_{i_2+1}}^\text{2-loop}(i_1,i_2,i_3,i_4)
\tr(1\cdots (i_2-1)b)\tr(i_2+2\cdots n)
\\
&+if^{a_{i_2+1}a_1b}\frac{1}{\spa{i_2+1,1}}
A_{n-1;c_{i_2+1}}^\text{2-loop}(i_1,i_2,i_3,i_4)
\tr(b2\cdots i_2)\tr(i_2+2\cdots n)
\\
&+if^{a_{i_2+1}a_{i_2+2}b}\frac{1}{\spa{i_2+1,i_2+2}}
A_{n-1;c_{i_2+1}}^\text{2-loop}(i_1,i_2,i_3,i_4)
\tr(1\cdots i_2)\tr(b(i_2+3)\cdots n)
\\
&+if^{a_na_{i_2+1}b}\frac{1}{\spa{n,i_2+1}}
A_{n-1;c_{i_2+1}}^\text{2-loop}(i_1,i_2,i_3,i_4)
\tr(1\cdots i_2)\tr(i_2+2\cdots (n-1)b)
\\
=& \, A_{n-1;c_{i_2+1}}^\text{2-loop}(i_1,i_2,i_3,i_4)
\frac{\spa{i_21}}{\spa{i_2,i_2+1}\spa{i_2+1,1}}
\tr(1\cdots i_2+1)\tr(i_2+2\cdots n)
\\
&+A_{n-1;c_{i_2+1}}^\text{2-loop}(i_1,i_2,i_3,i_4)
\frac{\spa{n,i_2+2}}{\spa{i_2+1,i_2+2}\spa{n,i_2+1}}
\tr(1\cdots i_2)\tr(i_2+1\cdots n)
\\
=&\, A_{n;i_2+2}^\text{2-loop}(i_1,i_2,i_3,i_4)
\tr(1\cdots i_2+1)\tr(i_2+2\cdots n)
\\
&+A_{n;i_2+1}^\text{2-loop}(i_1,i_2,i_3,i_4)
\tr(1\cdots i_2)\tr(i_2+1\cdots n).
\end{split}
\end{align}
The first equality follows from taking only the double-traces with the
ordering $1,\dotsc,n$ after removing the index contraction.
The second equality follows from the definition~\eqref{dtterm}.
This exhausts all cases, and the result follows by induction.

\section{Proof of eq.~\eqref{AGR}}
\label{sec:AGRproof}
	
In this section, we prove that the single-trace color-ordered one-loop
subamplitude when matter lives in the representation \eqref{therep}
is given by eq.~\eqref{AGR}, which we repeat here for convenience:
\be
-8A_n^{[1]}(1,\dotsc,n)
+ \sum_{k=1}^n\sum_{\sigma\,\in\,\alpha_k\shuffle\beta_k}A_n^{[1]}(1,\sigma),
\label{AGRreprise}
\ee
where $\alpha_k = (2,\dotsc,k)$ and $\beta_k=(k+1,\dotsc,n)$.
The first term immediately follows from the eight copies
of the fundamental representation for the fermions, together with the
sign flip associated with the SWI~\eqref{1loopsusy}.  
The second term must then come from the single copy of the antisymmetric
tensor representation, in particular from the exchange graph shown in
figure \ref{fig:antisym}.
(The gluon loop only generates double traces, and single traces with a
factor of $N$ which cancel against non-exchange contributions
from $\wedge^2 F$.)
Figure \ref{fig:shufflex} provides an example, for $n=6$,
of how a particular shuffle of $\alpha_4$ and $\beta_4$, i.e.~an element of
$\alpha_k\shuffle\beta_k$ for $k=4$, can contribute to the trace
ordering $\tr(1\cdots n)$.

\begin{figure}
\centering
\be
\nonumber
	\begin{tikzpicture}[baseline={(0,0)},scale=0.75]
		\begin{feynman}
			\vertex [label=left:$1$] (g1) at (0,-2);
			\vertex [label=left:$2$] (g2) at (0,2);
			\vertex [label=right:$3$] (g3) at (4,2);
			\vertex [label=right:$4$] (g4) at (4,-2);
			\vertex (v1) at (1,-1);
			\vertex (v2) at (1,1);
			\vertex (v3) at (3,1);
			\vertex (v4) at (3,-1);
			\vertex (w1) at (1.25,-0.75);
			\vertex (w2) at (1.25,0.75);
			\vertex (w3) at (2.75,0.75);
			\vertex (w4) at (2.75,-0.75);
			\vertex (b1) at (2+1.4*0.34202,-1.4*0.93969);
			\vertex (b2) at (2-1.4*0.34202,-1.4*0.93969);
			\vertex (b3) at (2.1,-1.049161396);
			\vertex (b4) at (1.9,-1.141211056);
                        \vertex (v5) at (0.95,0);
                        \vertex (v6) at (3.05,0);
                        \vertex [label=left:$5$] (g5) at (-0.5,0);
                        \vertex [label=right:$6$] (g6) at (4.5,0);
			\diagram*{
				(g1) -- [gluon] (v1),
				(v2) -- [gluon] (g2),
				(g3) -- [gluon] (v3),
				(v4) -- [gluon] (g4),
				(v1) -- [quarter left] (v2) 
				-- [quarter left, fermion] (v3) 
				-- [quarter left] (v4),
				(w1) -- [quarter left] (w2) 
				-- [quarter left, fermion] (w3) 
				-- [quarter left] (w4),
				(b1) -- (w1),
				(w4) -- (b3),
				(b4) -- (b2),
				(g5) -- [gluon] (v5),
				(v6) -- [gluon] (g6),
			};
			\draw ([shift=(-20:1.4)] 2, 0) arc (-20:-70:1.4);
			\draw ([shift=(200:1.4)] 2, 0) arc (200:250:1.4);
\end{feynman}
\end{tikzpicture}
\ee
\caption{An example for $n=6$ which illustrates
how the color-ordered sub-amplitude
$A_6^{[1]}(1,5,2,3,6,4)$ can contribute to the color factor $\tr(123456)$
via the exchange term $P$.
The ordering $(1,5,2,3,6,4)$ corresponds to a shuffle of
$\alpha_4 = (2,3,4)$ and $\beta_4 = (5,6)$.}
\label{fig:shufflex}
\end{figure}
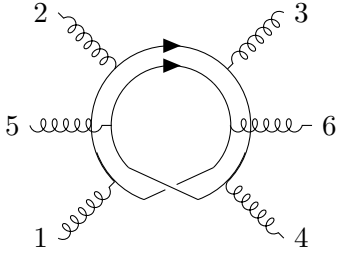
	
We now provide a rigorous argument for the validity of eq.~\eqref{AGR}.
That is, we will show that the contributing color-orderings
for the exchange terms are in
bijection with the shuffles $\alpha_k\shuffle\beta_k$ for some $k$.
Recall that the single-trace terms of the one-loop amplitude with matter
in the representation \eqref{therep} is given by
\begin{align}
\begin{split}
\sum_{\sigma\in S_n/\mathbb{Z}_n}
&~\Bigg[
-8\tr(\sigma_1\cdots \sigma_n)A_n^{[1]}(\sigma_1,\dotsc,\sigma_n)
\\
&+\frac{1}{2}\sum_{I\subset(1,\dotsc,n)}
\tr(\sigma(I\cdot I^c))A_n^{[1]}(\sigma_1,\dotsc,\sigma_n)
\Bigg] \,.
\end{split}
\end{align}
The sum is over the group $S_n/\mathbb{Z}_n\cong S_{n-1}$, allowing us
to choose an element from $1,\dotsc,n$ which can be fixed by all
$\sigma\in S_{n-1}$. We choose $1$ to be this fixed element.
After summing over $S_{n-1}$ and collecting on the traces,
the color-ordered term multiplying $\tr(1\cdots n)$ is generically
\be
-8A_n^{[1]}(1,\dotsc,n) + \frac{1}{2}\sum_{\sigma\in S}A_n^{[1]}(1,\sigma),
\ee
where $S\subset S_{n-1}$ is
\be
S = \{\sigma\in S_{n-1}|\sigma(I\cdot I^c)
\in [(1,\dotsc,n)]~\text{for some}~I\subset(1,\dotsc,n)\},
\ee
where is $[(1,\dotsc,n)]$ is an equivalence class containing all cycles of
$(1,\dotsc,n)$. Here $S$ is written as a set, but we are
counting multiplicities, meaning that $\sigma$ is included in the sum
the same number of times there is an instance of a sublist
$I\subset(1,\dotsc,n)$ with $\sigma(I\cdot I^c)\in[(1,\dotsc,n)]$. 
	
The set $S$ can be written as a disjoint union over subsets $\hat{S}_k$
which require the sublist $I$ to be of size $k$, allowing for the sum
over $S$ to be written as a sum over $k$,
\be
\frac{1}{2}\sum_{\sigma\in S}A_n^{[1]}(1,\sigma)
= \frac{1}{2}\sum_{k=0}^{n}\sum_{\sigma\in \hat{S}_k} A_n^{[1]}(1,\sigma),
\ee
where the collection of permutations $\hat{S}_k$ is
\be
\hat{S}_k = \{\sigma\in S_{n-1}|\sigma(I\cdot I^c)
\in [(1,\dotsc,n)]~\text{for some}~I\subset(1,\dotsc,n)~\text{with}~|I|=k\}.
\ee
Notice that if $\sigma\in \hat{S}_k$ then $\sigma\in \hat{S}_{n-k}$,
which follows from the fact that if $\sigma(I\cdot I^c)\in [(1,\dotsc,n)]$
then $\sigma(I^c\cdot I)\in [(1,\dotsc,n)]$, since $I^c\cdot I$ is related to
$I\cdot I^c$ by a cyclic transformation.
We can use this pairing between $I$ and $I^c$ to require that
$1\in I\subset(1,\dotsc,n)$.  This restriction removes the overall
factor of $1/2$ and the sum becomes
\be
\label{SktotildeSk}
\frac{1}{2}\sum_{k=0}^{n}\sum_{\sigma\in \hat{S}_k\subset S_{n-1}}A_n^{[1]}(1,\sigma)
= \sum_{k=1}^{n}\sum_{\sigma\in \tilde{S}_k}A_n^{[1]}(1,\sigma),
\ee
where the new subset $\tilde{S}_k$ is
\be
\tilde{S}_k = \{\sigma\in S_{n-1}|\sigma(I\cdot I^c)
\in[(1,\dotsc,n)]~\text{for some}~I\subset(1,\dotsc,n)
~\text{with}~|I|=k~\text{and}~1\in I\}
\ee
for all $1\leq k\leq n$. 
	
We will show that $\tilde{S}_k=\alpha_k\shuffle\beta_k$,
which will complete the proof of eq.~\eqref{AGR}.
As a reminder, $\alpha_k=(2,\dotsc,k)$ and $\beta_k=(k+1,\dotsc, n)$.
Consider an element $\tau\in \alpha_k\shuffle\beta_k$,
and set $J=(1,\tau^{-1}(\alpha_k))$.
The permutation $\tau$ is generically of the form 
\be
\tau=(\beta_{I_1},2,\beta_{I_2},3,\dotsc,\beta_{I_{k-1}},k,\beta_{I_k}),
\ee
where $\beta_{I_j}$ represents some sublist of $\beta_k$ such that
$\beta_{I_1}\cdot\beta_{I_2}\cdots\beta_{I_k}=\beta_k$.
Since $\tau\in S_{n-1}$, we can identify $\tau$ with $(1,\tau)\in S_n$.
So the $j$-th element of $\tau$ is in the $(j+1)$-th position in $(1,\tau)$.
Letting $j_i$ be the position of $i$ in $\tau$ for $2\leq i \leq n$,
we then have that 
\be
J = (1,j_2+1,j_3+1,\dotsc,j_k+1).
\ee
Also, $j_i<j_l$ for $i<l$, since $(\alpha_k)_i=i+1<l+1=(\alpha_k)_l$ and
the shuffle product preserves the ordering of $\alpha_k$. This means that
$J$ is ordered with respect to $(1,\dotsc,n)$.
It follows that the complement of $J$ is 
\be
J^c = (j_{k+1}+1, \dotsc, j_n+1) = \tau^{-1}(\beta_k).
\ee
Thus,
\be
\tau(J\cdot J^c)=(1,\alpha_k,\beta_k)=(1,2,\dotsc,n),
\ee
which implies that $\tau\in\tilde{S}_k$,
i.e.~$\alpha_k\shuffle\beta_k\subseteq\tilde{S}_k$.
	
The shuffle product $\alpha_k\shuffle\beta_k$ has size
\be
|\alpha_k\shuffle\beta_k| = \binom{|\alpha_k|+|\beta_k|}{|\alpha_k|}
=\binom{n-1}{k-1}.
\label{countshuffles}
\ee
The size of $\tilde{S}_k$ is at most the number of size-$k$ sublists
of $(1,\dotsc,n)$ containing $1$, i.e.
\be
|\tilde{S}_k|\leq\binom{n-1}{k-1}.
\label{boundtSk}
\ee
It cannot be larger, because $\sigma\in\tilde{S}_k$ if and only if there
exists $I\subset(1,\dotsc,n)$ containing $1$ such that
$\sigma(I\cdot I^c)\in[(1,\dotsc,n)]$. Since $\sigma\in S_{n-1}$ has $1$
as a fixed point, it must be that $\sigma(I\cdot I^c)=(1,\dotsc,n)$.
By uniqueness, this means there is only one such $\sigma$ for a given $I$.
So the size of $\tilde{S}_k$ is bounded by eq.~\eqref{boundtSk}.
Given that $\alpha_k\shuffle\beta_k\subseteq\tilde{S}_k$,
and eq.~\eqref{countshuffles}, the bound must be saturated,
and then $\tilde{S}_k = \alpha_k\shuffle\beta_k$ follows.
This proves the equality.

\section{Integrals}
\label{sec:integrals}

In this section, we reproduce the evaluated integrals from
ref.~\cite{Bern:2000dn} that enter the two-loop primitive amplitudes
in eqs.~\eqref{gluprimamps} and \eqref{primamps}. The results for the
two-loop integrals are given in the Euclidean region $s,t<0$ and $u>0$,
for which $\chi=t/s>0$. They can be analytically continued to other
regions by substituting 
$(-s)^{-\eps}\mapsto s^{-\eps}e^{\eps i\pi}$ and 
$\ln\chi\mapsto \ln|\chi|+i\pi$.
The planar double-box integral, expressed in terms of the one-loop box
integral, is
\begin{align}
\label{dimregPint}
\begin{split}
\mathcal{I}_4^P\big[\lam_p^2\lam_{p+q}^2\big](s,t)&=
\mathcal{I}_4^P\big[\lam_q^2\lam_{p+q}^2\big](s,t)
\\
&=-ic_\Gamma\frac{1}{\eps^2}(-s)^{-1-\eps} \, 
\mathcal{I}_4^\text{1-loop}[\lam_p^4](s,t) 
+ \frac{FR^P_{p+q,q}}{(4\pi)^4(-s)} + {\cal O}(\eps).
\end{split}
\end{align}
The one-loop box integral to ${\cal O}(\eps^2)$ is
\begin{multline}
\mathcal{I}_4^\text{1-loop}[\lam_p^4](s,t) = ic_\Gamma(-s)^{-\eps}(-\eps)(1-\eps)
\frac{1}{6}
\Bigg\{
\frac{1}{\eps} - \frac{1}{2}\frac{\chi(\ln^2\chi+\pi^2)}{(1+\chi)^2}
- \frac{\chi\ln\chi}{1+\chi} + \frac{11}{3}
\\
+\eps\Bigg[
\frac{\chi}{(1+\chi)^2}\bigg[
\Li_3(-\chi) - \zeta_3 - \ln\chi\Li_2(-\chi) + \frac{1}{3}\ln^3\chi 
- \frac{1}{2}\ln^2\chi\ln(1+\chi)
\\
+\frac{\pi^2}{2}\ln\Big(\frac{\chi}{1+\chi}\Big) 
+ \frac{1}{2}\Big( (2+\chi)\ln^2\chi + \pi^2 \Big)
\bigg]
\\
+\frac{11}{3}\bigg(
-\frac{1}{2}\frac{\chi(\ln^2\chi+\pi^2)}{(1+\chi)^2}
- \frac{\chi\ln\chi}{1+\chi} + \frac{11}{3}
\bigg)
-4
\Bigg]
\Bigg\}
+ {\cal O}(\eps^3).
\end{multline}
The planar finite remainder $FR^P_{p+q,q}$ is
\be
FR^P_{p+q,q} = \frac{1}{18}\frac{\chi}{(1+\chi)^2}
\bigg[
-\ln\chi(\ln^2\chi+\pi^2) + \bigg(\chi-\frac{1}{\chi}\bigg)\pi^2
\bigg].
\ee

The divergent non-planar integral in terms of the one-loop box integral is
\be
\label{dimregNPint}
\mathcal{I}_4^{NP}[\lam_p^2\lam_q^2](s,t)
= -ic_\Gamma\frac{1}{\eps^2}(-s)^{-1-\eps} \,
\mathcal{I}_4^\text{1-loop}[\lam_p^4](u,t)
+ \frac{FR_{p,q}^{NP}}{(4\pi)^4(-s)},
\ee
where the finite remainder is
\begin{multline}
FR^{NP}_{p,q} = {1\over6} \Biggl\{
 - 2 \chi(1+\chi) \biggl[
         \Li_3\Bigl({\chi\over1+\chi}\Bigr) - \zeta_3
       - \ln\Bigl({\chi\over1+\chi}\Bigr)
          \Bigl( \Li_2\Bigl({\chi\over1+\chi}\Bigr) + {\pi^2\over2} \Bigr)
       - {1\over6} \ln^3\Bigl({\chi\over1+\chi}\Bigr) \biggr]
\\
 + 3 \chi(1+\chi) \ln(1+\chi) \ln\chi
 - {1\over2} (1+\chi)^2 \biggl( -{1\over\chi} + 3 \biggr) \ln^2(1+\chi)
 - {1\over2} \chi^2 \biggl( {1\over1+\chi} + 3 \biggr) \ln^2\chi
\\
 + \pi^2 \biggl( \chi - {1\over2} {1\over 1+\chi}
              + {5\over6} \biggr)
 + (1+\chi) \ln(1+\chi) - \chi \ln\chi
\\
+ i \pi \biggl(
    2 \chi(1+\chi) \biggl[ \Li_2\Bigl({\chi\over1+\chi}\Bigr)
                            - {\pi^2\over6} - {3\over2} \ln\chi  \biggr]
  + (1+\chi) \biggl[
         (1+\chi) \Bigl( -{1\over\chi} + 3 \Bigr) \ln(1+\chi) - 1 \biggr]
  \biggr) \Biggr\}
\,.
\end{multline}

The finite non-planar integral 
$\mathcal{I}_4^{NP}[\lam_q^2\lam_{p+q}^2]=\mathcal{I}_4^{NP}[\lam_p^2\lam_{p+q}^2]$
is
\begin{multline}
\mathcal{I}_4^{NP}[\lam_q^2\lam_{p+q}^2](s,t)  =  \frac{1}{ (4 \pi)^4(-s)} 
\frac{1}{6}  \Biggl\{ 
{\chi \over (1+\chi)^2} \biggl[  \Li_3(-\chi) - \zeta_3 
- \ln\chi \Bigl( \Li_2(-\chi) - \frac{\pi^2}{6} \Bigr)
- \frac{3}{4} \chi \bigl( \ln^2\chi - \pi^2 \bigr) \biggr]
\\
- \frac{1+\chi}{\chi^2} \biggl[  \Li_3\Bigl({1\over1+\chi}\Bigr) - \zeta_3 
+ \ln(1+\chi) \biggl( \Li_2\Bigl({1\over1+\chi}\Bigr) 
+ {\pi^2\over6} \biggr)
\\
+ {3\over4} (1+\chi) \ln^2(1+\chi)
+ {1\over3} \ln^3(1+\chi) \biggr]
+ \biggl( {1\over \chi(1+\chi)} + {3\over2} \biggr) 
\ln(1+\chi) \ln\chi
\\
+ \pi^2 \biggl( {1 \over 6\chi} + {4 \over 3(1+\chi)}
+ {3\over2} {\chi\over(1+\chi)^2} - {3\over4} \biggr)
+ {\ln(1+\chi)\over 2\chi} - {\ln\chi\over 2(1+\chi)} 
\\
+ i \pi \biggl( 
- { 1+\chi \over \chi^2 } \biggl[ 
\Li_2\Bigl( { \chi \over 1+\chi } \Bigr)
- \ln(1+\chi) \ln\chi  
+ {1\over2} \ln^2(1+\chi)  
- {3\over2} (1+\chi) \ln(1+\chi) \biggr]
\\
- {1\over2} {\chi\over(1+\chi)^2} 
\bigl( \ln^2\chi + \pi^2 \bigr)
- \biggl( {1 \over \chi(1+\chi)} + {3\over2} \biggr) \ln\chi
- {1\over2\,\chi}\ \biggr)  \Biggr\}.
\end{multline}
	
Finally, we provide parts of the explicit expressions for
eqs.~\eqref{A411func} and \eqref{A430func}. They are
\begin{multline}
A_{4;1;1}(1,2,3,4)
= \frac{4}{3}\rho c_\Gamma^2\frac{(-s)^{-2\epsilon}}{\chi^2(1+\chi)^2}
\frac{1}{\epsilon}
\\
\times\Bigg[
\chi^3(3+3\chi+3\chi^2+\chi^3)\ln^2\bigg(\frac{\chi}{1+\chi}\bigg)
+4\chi^3\ln(1+\chi)\ln\bigg(\frac{\chi}{1+\chi}\bigg)
\\
+(1+3\chi+3\chi^2+3\chi^3)\ln^2(1+\chi)
+2\pi^2\chi^3
\\
+2\chi^3(1+\chi)(3+\chi)\ln\bigg(\frac{\chi}{1+\chi}\bigg)
-2\chi(1+\chi)(1+3\chi)\ln(1+\chi)
\\
+ 2i\pi(1+\chi)^3
\bigg\{
\chi^3\ln\bigg(\frac{\chi}{1+\chi}\bigg)-\ln(1+\chi)+\chi(1+\chi)
\bigg\} 
\Bigg]
+ {\cal O}(\eps^0)
\end{multline}
and 
\begin{multline}
\label{A430eq}
A_{4;3;0}(1,2,3,4)
= \frac{4}{3}\frac{\rho}{(4\pi)^4}\frac{1}{\chi^2(1+\chi)^2}
\\
\times \Bigg\{
\chi^4(3+3\chi+\chi^2)\ln^3\bigg(\frac{\chi}{1+\chi}\bigg)
-(1+3\chi+3\chi^2)\ln^3(1+\chi)
\\
+\chi^3(1+6\chi+6\chi^2+2\chi^3)
\bigg(
\ln(1+\chi)\ln\bigg(\frac{\chi}{1+\chi}\bigg) + \pi^2 \bigg)
\ln\bigg(\frac{\chi}{1+\chi}\bigg)
\\
-(2+6\chi+6\chi^2+\chi^3)
\bigg(
\ln(1+\chi)\ln\bigg(\frac{\chi}{1+\chi}\bigg)+\pi^2 \bigg)
\ln(1+\chi)
\\
+2\chi^4(1+\chi)\ln^2\bigg(\frac{\chi}{1+\chi}\bigg)
+2\chi(1+\chi)\ln^2(1+\chi)
\\
+2\chi(2+\chi+2\chi^2)(1+\chi)^2\ln(1+\chi)\ln\bigg(\frac{\chi}{1+\chi}\bigg)
\\
+ 2\pi^2\chi(1+\chi)^4 + 18\chi^2(1+\chi)^2
\\
+i\pi\bigg[
4(1+3\chi+3\chi^2+\chi^3+3\chi^4+3\chi^5+\chi^6)\ln(1+\chi)
\ln\bigg(\frac{\chi}{1+\chi}\bigg)
\\
+\chi^3(-1+3\chi+3\chi^2+\chi^3)\ln^2\bigg(\frac{\chi}{1+\chi}\bigg)
+(1+3\chi+3\chi^2-\chi^3)\ln^2(1+\chi)
\\
-2\chi(1+\chi)(2+3\chi+3\chi^2)\ln\bigg(\frac{\chi}{1+\chi}\bigg)
+2\chi^2(1+\chi)(3+3\chi+2\chi^2)\ln(1+\chi)
- 2\pi^2\chi^3
\bigg]
\Bigg\} + {\cal O}(\eps).
\end{multline}
We only give the lowest order in $\eps$ term for $A_{4;1;1}$ due to the
complexity of the ${\cal O}(\eps^0)$ term. It is already evident at this order
that the dimensionally-regulated YM amplitude does not agree with
eq.~\eqref{KC4pt}, the sdYM form-factor result.  The predicted
answer from the CCA bootstrap for $A_{4;3;0}$
is merely the $18\chi^2(1+\chi^2)$ term in eq.~\eqref{A430eq}.
	
\bibliographystyle{JHEP}
\bibliography{gaugeamps2.bib}
		
\end{document}